\documentclass[3p,number,sort&compress,twocolumn]{elsarticle}

\usepackage{graphics}
\usepackage{graphicx}
\usepackage{epsfig}
\usepackage{amssymb}
\journal{JMMM}

\def\bscco{Bi$_2$Sr$_2$CaCu$_2$O$_{8+\delta}$}

\def\lco{La$_2$CuO$_4$}

\def\lsco{La$_{2-x}$Sr$_x$CuO$_4$}
\def\lbco{La$_{2-x}$Ba$_x$CuO$_4$}

\def\lbcoate{La$_{1.875}$Ba$_{0.125}$CuO$_4$}

\def\ybco{YBa$_2$Cu$_3$O$_{6+x}$}

\def\bscco{Bi$_2$Sr$_2$CaCu$_2$O$_{8+\delta}$}

\begin{document}

\begin{frontmatter}


\title{Superconductivity, Antiferromagnetism, and Neutron Scattering}
%\title{Neutron Scattering Studies of High Temperature Superconductors}
% \author{Author's Name\corauthref{cor1}\thanksref{label2}}
\author{John M. Tranquada,$^*$ Guangyong Xu, and Igor A. Zaliznyak}
\address{Condensed Matter Physics \&\ Materials Science Dept., Brookhaven National Laboratory, Upton, NY 11973-5000, USA}
 %\address{Address\thanksref{label3}}
 \cortext[cor1]{Corresponding author, jtran@bnl.gov}

%\title{}


%\author{}

%\address{}

\begin{abstract}
High-temperature superconductivity in both the copper-oxide and the iron-pnictide/chalcogenide systems occurs in close proximity to antiferromagnetically ordered states.  Neutron scattering has been an essential technique for characterizing the spin correlations in the antiferromagnetic phases and for demonstrating how the spin fluctuations persist in the superconductors.  While the nature of the spin correlations in the superconductors remains controversial, the neutron scattering measurements of magnetic excitations over broad ranges of energy and momentum transfer provide important constraints on the theoretical options.  We present an overview of the neutron scattering work on high-temperature superconductors and discuss some of the outstanding issues.
\end{abstract}

%\begin{keyword}
%     high-temperature superconductors   \sep copper oxides \sep stripes
%\PACS 74.72.-h \sep 75.25.Dk   \sep 74.81.-g
%\end{keyword}
\end{frontmatter}

% main text
\section{Introduction}
\label{}

Looking at temperature vs.\ composition phase diagrams of high-temperature superconductors, including both cuprate and iron-pnictide/chalcognide systems, one finds antiferromagnetic (AF) order in close proximity to superconductivity \cite{mazi10,lums10r,scal12a}.   This close association, together with experimental evidence for AF correlations in superconducting samples, has led many theorists to a common belief that AF fluctuations play an important role in the electron-pairing mechanism that underlies superconductivity \cite{lee06,kive07,scal12a}.  The differences among theoretical perspectives only begin to appear when one considers the specifics of how the AF correlations impact pairing.  To understand the source of these disagreements, one must step back and recognize that there is no consensus on how to describe the interaction between charge carriers and spin excitations in a metallic conducting system with strong AF correlations, especially when the system is close to a transition to a correlated (Mott) insulator state.  In the absence of a common view on how to frame the problem, it should not be surprising that there is a lack of consensus on the approach to a solution.

Neutron scattering is the preeminent technique for studying AF spin fluctuations in solids.  One can, of course, also obtain important information on local hyperfine fields, susceptibilities, and slow fluctuations from techniques such as nuclear magnetic resonance (NMR) and muon spin rotation, and magnetic order can be probed with resonant x-ray scattering.  The defining feature of AF fluctuations in high-temperature superconductors, however, is their remarkably high energy scale.  Magnetic excitations in these systems extend up to energies of several hundreds of meV, which easily exceeds the maximum energy of phonon excitations involved in the traditional mechanisms of superconductivity. This observation makes highly energetic  AF fluctuations a primary suspect for mediating the unconventional high-temperature superconductivity.

To characterize the AF correlations in these and other strongly-correlated metallic systems, it is therefore crucial to cover the energy range from fractions to hundreds of meV and to probe all of reciprocal space.  Developments in spectrometers and sources over the last two decades have greatly improved the efficiency of such measurements.  The throughput of triple-axis spectrometers has been enhanced through the use of multiple analyzers and detectors, beginning with developments such as RITA \cite{rita06} (first at Ris\o\ and continuing at SINQ) and SPINS \cite{zali05} and progressing to MACS \cite{macs08} and BT7 \cite{lynn12} at the NIST Center for Neutron Research; of course, enhanced instruments are available at many other facilities, including the Institut Laue Langevin, FRM-II, and the Laboratoire L\'eon Brillouin.  While most triple-axis instruments  are optimized for excitations in the cold to thermal neutron range ($E\lesssim30$~meV) 
time-of-flight spectrometers at spallation sources have enabled measurements with epithermal neutrons probing excitations up to $\sim1$~eV, and covering a very large phase space.  This began with the seminal MAPS spectrometer \cite{maps94} (and later MERLIN \cite{merlin06}) at ISIS, and has become common place with instruments such as ARCS \cite{arcs12} and SEQUOIA \cite{sequoia10} at the Spallation Neutron Source (SNS) and 4SEASONS \cite{4seasons11} at the Japanese Proton Accelerator Research Complex.  Very soon, the SNS spectrometer HYSPEC will have the capability of distinguishing magnetic from nuclear scattering through neutron-spin polarization analysis \cite{hyspec06}, a technique that has long been used to advantage on triple-axis instruments \cite{regn03}.

Ideally, experimentalists should be able to interpret their results by comparing with existing theoretical expressions or to simply measure the magnetic response in a superconductor family as a function of doping and temperature, and then let the theorists interpret the results.  In practice, however, things are not so simple.  The neutron scattering cross section is relatively weak, so that large crystals ($\gtrsim 1$~cm$^3$) or assemblies of crystals are typically required for each composition to be studied.  In some cases, it has taken decades for crystal growth technology and know-how to evolve to the point that suitable crystals are available.  Another aspect involves interpretation of the measurements. Frequently, there is a disconnect between theoretical interpretations and experimental results arising from various assumptions and idealizations adopted both in analyzing the data and applying the theoretical model. To present results in a useful fashion, it is generally necessary to parametrize the data by fitting with a model.  The models used, or the words used to describe them, are often based on specific theoretical perspectives.  Without a general and accepted theoretical description of itinerant antiferromagnetism, there is no universally accepted and unbiased language for describing measurements of spin fluctuations in high-temperature superconductors.

In the following sections, we will briefly summarize what we consider to be the important results from neutron scattering studies on antiferromagnetism in hole-doped cuprates and in iron-based superconductors.  We will also point to some of the open questions.  Obviously, our choices and presentation reflect our own biases.  The coverage of experiments and theory is necessarily in%
complete, and we refer the interested reader to more extensive review articles on cuprates \cite{fuji12a,kast98,bour98,maso01,lynn01,tran07,kive03,deml04,lee06,esch06,ogat08,birg06} and iron-based systems \cite{lynn09,lums10r,pagl10,john10,wen11,dai12}.

\section{Notes on neutron scattering}

Neutron scattering measures the product of the dynamical spin structure factor ${\cal S}({\bf Q},\omega)$, which is the Fourier transform of the spin-spin correlation function, and the square of the magnetic form factor $F({\bf Q})$, which is the Fourier transform of the density of the electronic magnetization cloud associated with each spin, normalized to one at ${\bf Q} = 0$.  Here $\hbar{\bf Q}$ is the neutron momentum transfer and $-\hbar\omega$ is the energy transfer. The dynamical structure factor describes the cooperative behavior of electronic spin variables, including spin order and excitations, whereas the form factor relates it to the behavior of the magnetization density in the crystal. This latter is the quantity which interacts with neutron's magnetic moment and is probed in experiment. The magnetic form factor is determined by the Wannier functions of magnetic electrons, which could be obtained from first-principles calculations.  A prediction for the dynamical structure factor, ${\cal S}({\bf Q},\omega)$, can in principle be obtained from the theoretical analysis of the model spin Hamiltonian that describes the system.  For an ordered antiferromagnet, the excitations are spin waves that can be calculated in perturbation theory, where one assumes that the fluctuations are small compared to the ordered moment.  As we will discuss, the fluctuations become more important when one reduces the size of the spin and the dimensionality.  In the case of the weakly-coupled planes with spin $S=\frac12$ that occur in cuprates, spin-wave theory is not a well-controlled approximation, though it remains a useful description.

${\cal S}({\bf Q},\omega)$ obeys a number of useful sum rules, which are helpful in analyzing the experimental neutron scattering data \cite{zali05}. In particular, for a system of identical spins $S$, one has
\begin{equation}
  C\int_{-\infty}^{\infty}d\omega \int_{\rm BZ}d{\bf Q}\, {\cal S}({\bf Q},\omega) = S(S+1),
  \label{eq:sum}
\end{equation}
where $C=v_0/(2\pi)^3$ and $v_0$ is the unit cell volume.  It is possible to convert ${\cal S}({\bf Q},\omega)$ to the imaginary part of a generalized spin susceptibility with the formula
\begin{equation}
  \chi''({\bf Q},\omega) = \pi \left(1-e^{-\hbar\omega/kT}\right)
S({\bf Q},\omega).
\end{equation}
By integrating $\chi''({\bf Q},\omega)$ over {\bf Q}, one obtains the local susceptibility $\chi''(\omega)$.

\section{Cuprates}

\subsection{Parent insulators}

The parent materials of the cuprate superconductors are correlated insulators.  Taking \lco\ as an example, it has a charge excitation gap of $\sim2$~eV \cite{kast98}, so that the only low energy excitations involve spin fluctuations. These can be described by an effective spin Hamiltonian, where the nearest-neighbor Cu spins within the planes are coupled by a superexchange energy $J$ that is quite large, while the effective coupling between planes is extremely weak, so that there are strong two-dimensional (2D) AF spin correlations at temperatures far above magnetic ordering temperature $T_N$ \cite{birg99}.

The Cu moments within the CuO$_2$ planes order antiferromagnetically below a N\'eel temperature $T_N=325$~K in stoichiometric \lco\ \cite{keim92a}.  Neutron diffraction studies have shown that the ordered moment is quite small, $\approx 0.6\mu_B$ \cite{yama87}. This corresponds to an ordered spin value of $\langle S \rangle \approx 0.27$ (assuming $g \approx 2.2$ for the spectroscopic Lande $g-$factor of Cu \cite{huck08}), which is even smaller than the value of 0.30 predicted by spin-wave theory upon accounting for zero-point motion \cite{ande52,oguc60,mano91}.  Similar results have been obtained for other cuprate antiferromagnets \cite{tran07}. Consequently, the contribution of the static spin order, which is measured by elastic Bragg peaks (delta-functions in energy), $\langle S \rangle ^2$,  accounts for only $\sim10\%$ of the total spin spectral weight, $S(S+1) = 3/4$, in the sum rule for $S = 1/2$. Hence, more than 90\%\ of the spectrum is inelastic. 

\begin{figure}[t]
\centerline{\includegraphics[width=6cm]{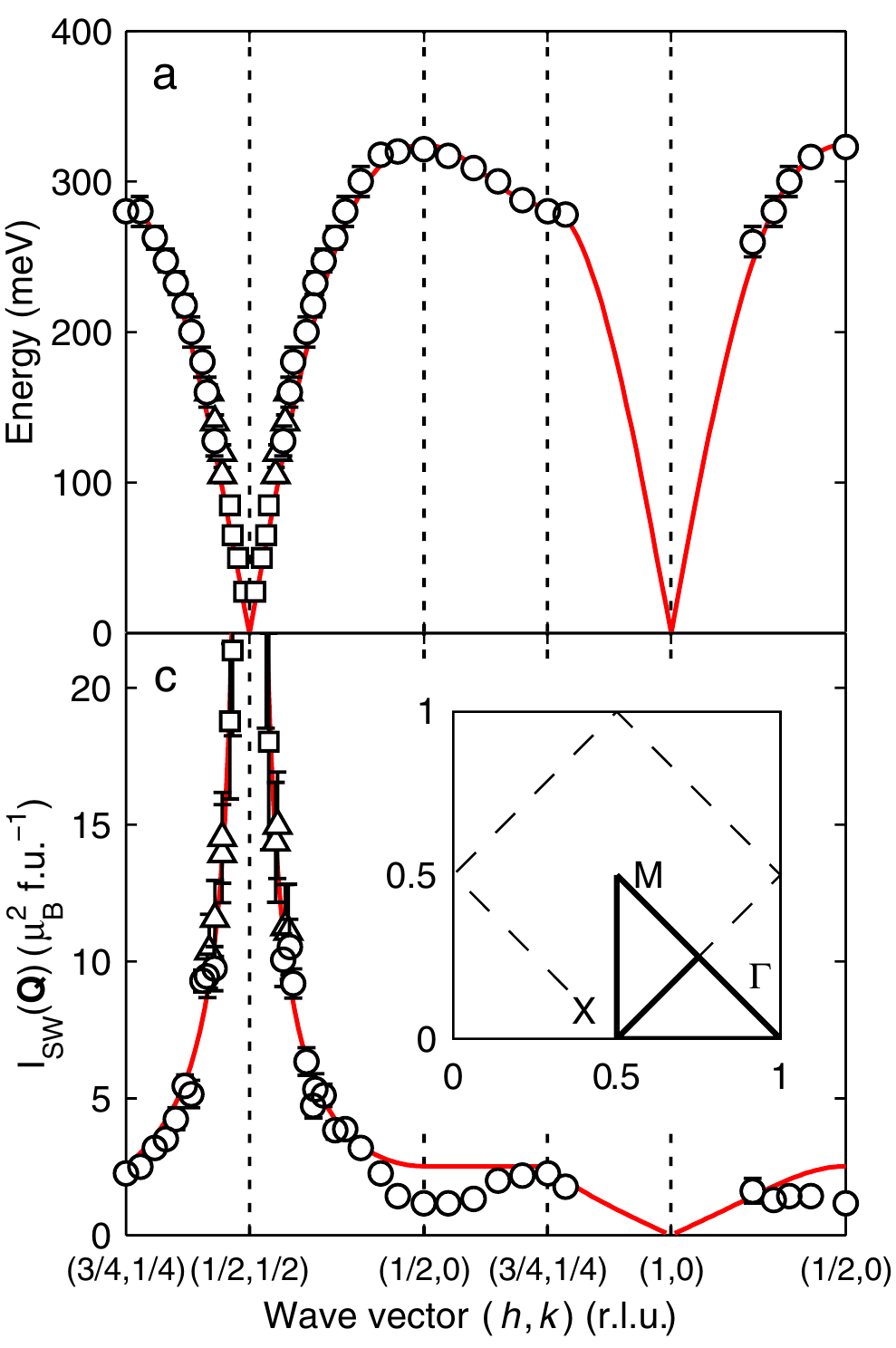}}
\caption{(a) Spin wave dispersion in La$_2$CuO$_4$ measured by inelastic neutron scattering, plotted for high-symmetry directions, as indicated in the inset of (c).  (c) Measured single magnon intensity vs.\ wave vector.  In both panels, the lines represent fits with a spin-wave model.   Reprinted figure with permission from Headings {\it et al.}\ \cite{head10}, \copyright\  2010 American Physical Society.}
\label{fg:lco}
\end{figure}

Figure~\ref{fg:lco} shows a recent measurement of the spin 
excitations in antiferromagnetic \lco\ by Headings {\it et al.} \cite{head10}.  
The authors find that both the dispersion and the wave-vector dependence of the intensity are described rather well by spin-wave theory expressions. The absolute intensity, however, appears much lower than predicted by the linear spin-wave theory, requiring a downward renormalization by a factor of $Z_d=0.4\pm0.04$. This is somewhat smaller than the value $Z_d\approx 0.6$ that is predicted from quantum corrections to linear spin waves \cite{lore05}. 

A fit to the dispersion in \lco\ yields $J=143\pm2$~meV, but also requires longer-range exchange couplings---second and third neighbor couplings $J'$ and $J''$, which are relatively weak, and a significant 4-spin cyclic exchange term $J_c\approx 0.4J$.  The overall bandwidth of the magnetic spectrum is $\sim2J$.  At the highest energies, there are some modest deviations in the spectral shape relative to the single-mode spectrum predicted by spin-wave theory. These seem to be consistent with corrections obtained in quantum Monte Carlo calculations \cite{sand01}. Similar findings were recently reported for several other cuprates by Dalla Piazza {\it et al.} \cite{dall12}, where the spin excitation spectra can be well described by a perturbative (up to a second order) treatment of an effective 1-band Hubbard model with nearest- and next-nearest neighbor hopping, or by a spin-wave treatment of the local spin Heisenberg Hamiltonian with extended-range interactions $J, J', J''$ and $J_c$.    Quantitative estimates for $J$ require that one take account of the bridging O atoms when evaluating the Cu to Cu hopping \cite{zaan87}.  The 4-spin cyclic exchange can be obtained at the same order of approximation in such a multi-orbital model \cite{roge89}.

Evaluating the sum rule for \lco\ without any model assumptions for ${\cal S}({\bf Q},\omega)$, the result is $\sim60$\%\ of the expected result for $S=\frac12$ \cite{head10}. Although part %Part
of the missing weight could be in multi-magnon excitations that are out of the measurement range, a significant missing weight may result from not taking proper account of the distribution of spin density.  Shamoto {\it et al.} \cite{sham93} have shown that accounting for the $3d_{x^2-y^2}$ orbital anisotropy of the Cu$^{2+}$ magnetic form factor is absolutely essential to understand the magnetic Bragg diffraction in YBa$_2$Cu$_3$O$_{6.15}$; it is also important for the analysis of the spin dynamics \cite{head10,cold01}. Although using the anisotropic ionic magnetic form factor of Cu$^{2+}$ is much better than using a spherical form factor of the dipole approximation, it still neglects the effect of covalency, {\it i.e.}\ charge transfer to the oxygen neigbors, which turns out to be very significant in the cuprates.

\begin{figure}[t]
\centerline{\includegraphics[width=6cm]{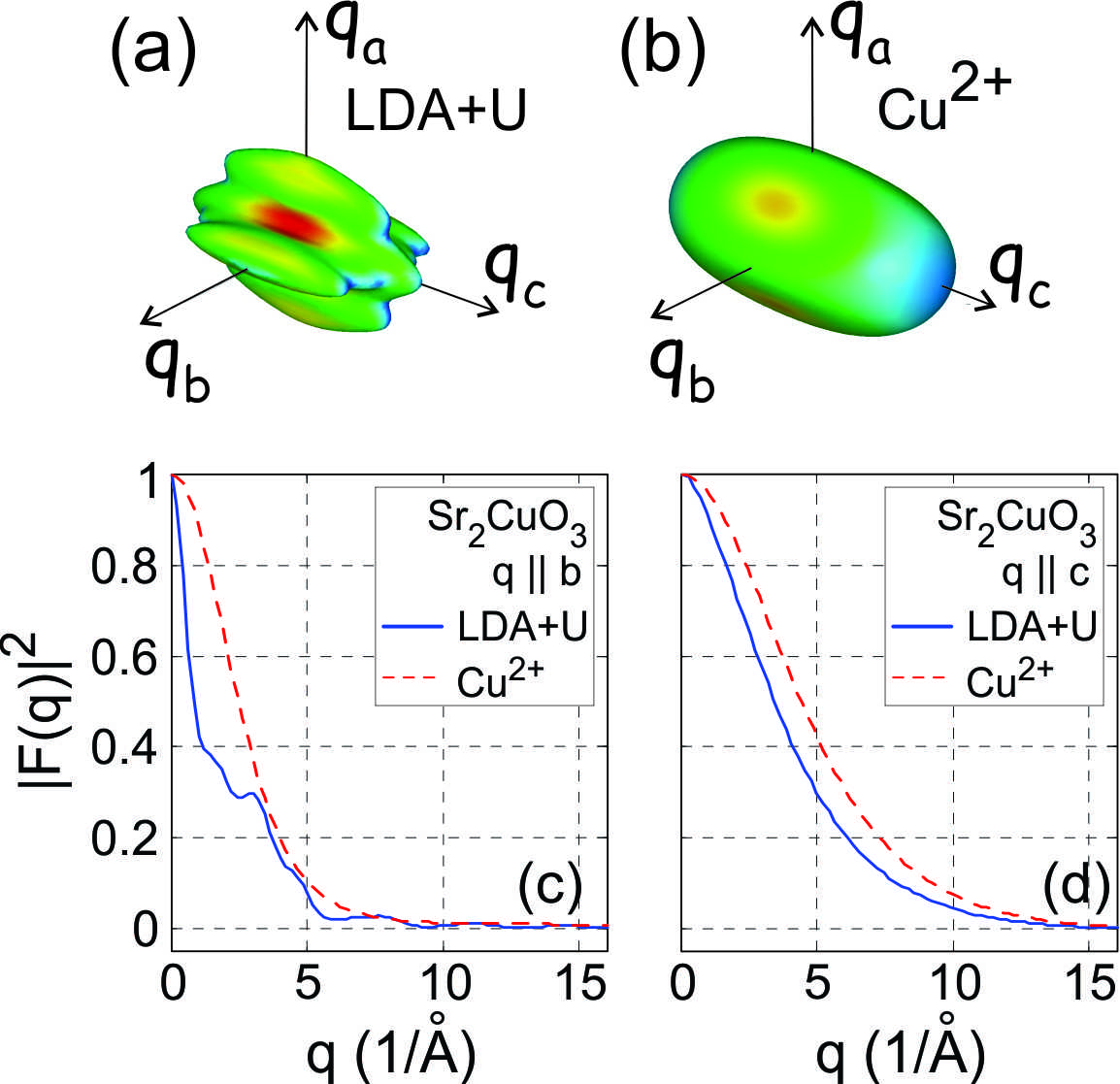}}
\caption{(a) Constant magnitude surface of the magnetic form $F({\bf Q})$ calculated from the Wannier function for a Cu $3d_{x^2-y^2}$ orbital hybridized with its neighbors in Sr$_2$CuO$_3$.  (b) Similar plot for an ionic Cu $3d_{x^2-y^2}$ state. (c), (d) Comparison of cuts of the magnetic form factors shown in (a) (solid blue line) and (b) (dashed red line) along two symmetry directions. From Walters {\it et al.}\ \cite{walt09}.}
\label{fg:ff}
\end{figure}

Walters {\it et al.}\ \cite{walt09} did a careful study of the form factor in the quasi-1D antiferromagnet Sr$_2$CuO$_3$. Local structure of the planar Cu--O square plaquettes in this material is essentially identical to that in \lco.  Making use of a precise theoretical result for the excitation spectrum, they demonstrated that a good fit to the data requires a form factor that takes account of hybridization between the half-filled Cu $3d_{x^2-y^2}$ orbital and the ligand O $2p_\sigma$ orbitals, as given by a density functional calculation.  The hybridization causes the spin density to be extended in real space, resulting in a more rapid fall off in reciprocal space compared to a simple Cu$^{2+}$ form factor, as illustrated in Fig.~\ref{fg:ff}. Smaller values of magnetic form factor at the wave vectors where the measurement is performed lead to the suppression of magnetic intensity, which could be as large as a factor of two or more \cite{walt09}. Finally, we note that a study of covalent NMR shifts in by Walstedt and Cheong \cite{wals01} found that barely 2/3 of the spin density in \lco\ resides on the copper sites, in excellent agreement with the Sr$_2$CuO$_3$ neutron data of Walters {\it et al.}\ \cite{walt09}.

\subsection{Hole-doped superconductors}

The antiferromagnetic insulator state is relatively well understood.  To get superconductivity, it is necessary to dope holes into the CuO$_2$ planes.  What happens to the spin fluctuations?

\begin{figure}[t]
\centerline{\includegraphics[width=5cm]{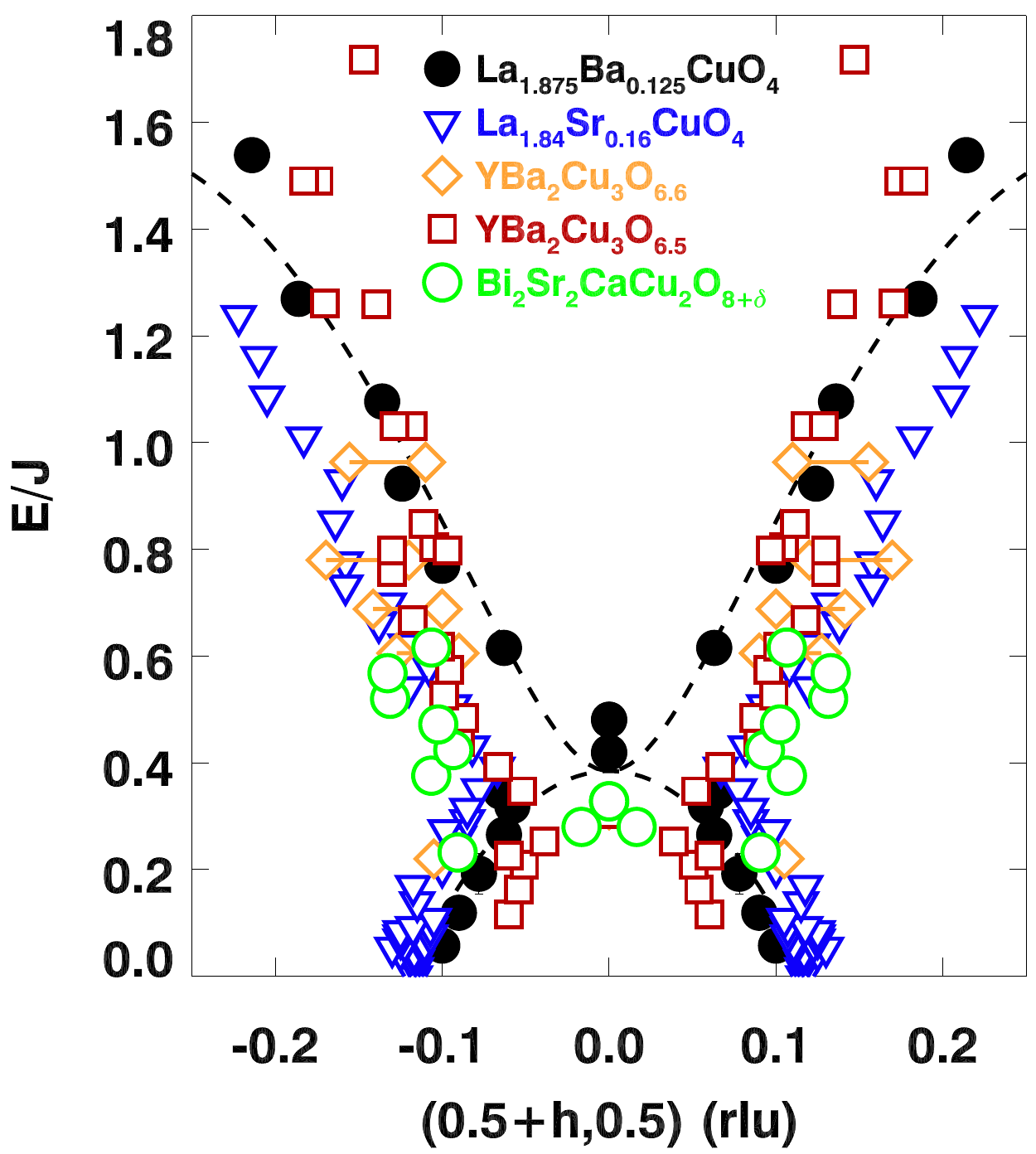}}
\caption{Magnetic dispersion relation along ${\bf Q}=(0.5+h, 0.5, 0)$ in various cuprates, corresponding to wave vectors parallel to the Cu-O bonds.  Data are for \lbcoate,\cite{tran04} La$_{1.84}$Sr$_{0.16}$CuO$_4$,\cite{vign07} YBa$_2$Cu$_3$O$_{6.6}$,\cite{hayd04} YBa$_2$Cu$_3$O$_{6.5}$,\cite{stoc05,stoc10} and \bscco.\cite{xu09} The energy is scaled by $J$ for the AF parent material.\cite{tran07,suga03}  From Fujita {\it et al.} \cite{fuji12a}.}
\label{fg:hourglass}
\end{figure}

Figure~\ref{fg:hourglass} summarizes the effective magnetic dispersion found in underdoped to optimally-doped cuprates.  The energy scale is normalized to $J$ of the undoped parent materials.  (Note that $J$ can vary by 30\%\ among different cuprate families \cite{bour97,suga03}.)  There are two important observations to make regarding this hour-glass-like spectrum: 1) the scale of the high-energy excitations seems to be $J$, characteristic of the correlated insulator state, and 2) the low-energy excitations are incommensurate.   Concerning point (1), the energy of the neck of the hour glass, $E_{\rm cross}$, decreases towards zero as one reduces the doping and approaches the antiferromagnetic state \cite{fuji12a}.  Also, as we will discuss shortly, the spectral weight of the magnetic excitations for $E\gtrsim E_{\rm cross}$ is observed to evolve smoothly from the AF state.

\begin{figure}[b]
\centerline{\includegraphics[width=8cm]{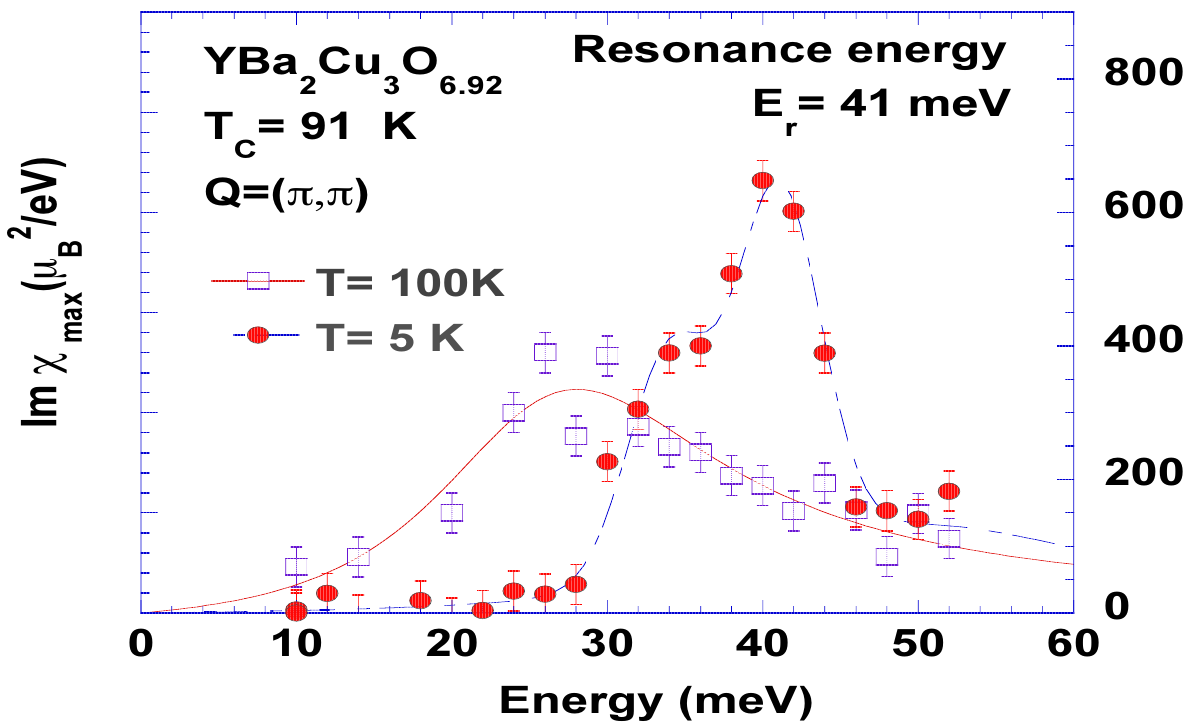}}
\caption{Measurements of $\chi''({\bf Q}_{\rm AF},\omega)$ below and above $T_c$ demonstrating the resonance and spin gap in YBa$_2$Cu$_3$O$_{6.92}$; from Bourges {\it et al.} \cite{bour99}.}
\label{fg:resonance}
\end{figure}

For $E\lesssim E_{\rm cross}$, samples near optimal doping and above develop a gap in the spin excitations and a resonance peak when the temperature drops below the superconducting transition, $T_c$.  An example of the change in the magnetic response at the AF wave vector for nearly-optimally-doped \ybco\ is shown in Fig.~\ref{fg:resonance}.  Here the magnetic signal is presented in terms of the imaginary part of the dynamical spin susceptibility,  $\chi''({\bf Q},\omega)$.
The energy of the resonance, $E_r$, varies with doping and among different cuprate families.  From measurements on \ybco\ and \bscco, Sidis {\it et al.} \cite{sidi04} found that $E_r\approx 5.3kT_c$.  Alternatively, Yu {\it et al.} \cite{yu09} compared with the energy scale $\Delta$ of the $d$-wave superconducting gap determined in photoemission studies and found $E_r/2\Delta\approx0.64$ for a large variety of materials.

How does one explain the temperature-dependent spin gap and resonance peak?  One way is to assume that the same electrons that go superconducting are also responsible for the magnetic response.  In this case, the magnetic response is presumably due to scattering the conduction electrons across the Fermi surface.  If the Fermi surface is ``nested'', meaning that it has parallel portions separated approximately by the AF wave vector, then one can obtain a substantial spin response \cite{viro90}, as occurs in the spin-density-wave state of metallic chromium \cite{fawc88}.   The opening of the superconducting gap removes electronic states that can contribute to $\chi''$, thus resulting in the spin gap.   The resonance peak is a consequence of BCS coherence factors plus a superconducting gap that changes sign between the nested portions of the Fermi surface \cite{scal12a}.  This is consistent with the $d$-wave gap of the cuprates \cite{tsue00}.

There have been many theoretical calculations of the resonance, as well as the downwardly dispersing excitations below it \cite{esch06}.  They can describe rather well the resonance in \ybco, which is commensurate with the AF wave vector, and they can provide some qualitative agreement with the downwardly-dispersing excitations; however, there are also significant discrepancies.  For $T\ll T_c$ and $\hbar\omega \ll 2\Delta$, the magnetic response should be determined by the positions of the nodes of the superconducting gap \cite{lu92}, but experimentally the low-energy incommensurate scattering is rotated 45$^\circ$ from the predicted direction \cite{hink10}, and there is no sign of the predicted signal even at the lowest energies \cite{lake99}.   Furthermore, the resonance in optimally and over-doped \lsco\ occurs at incommensurate wave vectors \cite{chri04,tran04}, rather than the commensurate position typically predicted by the calculations \cite{esch06}.

An alternative explanation for the low-energy incommensurate spin excitations attributes them to an electronically heterogeneous state involving alternating charge and spin stripes \cite{kive03,zaan01,vojt09}.  Static ordering of charge and spin stripes is observed in \lbco\ \cite{huck11} and related systems \cite{fuji12a} in which a subtle structural transition breaks the rotational symmetry of the planar Cu-O bonds, allowing a unique orientation of the stripes within each layer.  At low energies, spin fluctuations rise out of the incommensurate magnetic superlattice peaks, and at higher energies they exhibit the hour-glass spectrum of other cuprates, as shown for \lbco\ with $x=1/8$ in Fig.~\ref{fg:hourglass}.   Although stripe order tends to compete with 3D superconducting phase order, it coexists with 2D superconducting correlations \cite{li07,tran08} and even layered phase-decoupled superconductivity \cite{steg12}.   It has been proposed that the coexisting spin and superconducting orders are intertwined in a pair-density-wave (PDW) state \cite{berg09b}.  In such a state, the Cu moments providing the spin response are distinct from the doped holes that form the PDW superfluid.

It is worth noting that at higher temperatures, where the stripe order is lost, the spin excitations remain incommensurate \cite{fuji04,xu07} and are quite similar to those in optimally-doped \lsco\ \cite{aepp97}.  The anisotropic bulk susceptibility in paramagnetic \lbco\ is consistent with local moment behavior \cite{huck08}.  In \ybco, the modulation of the magnetic response as a function of momentum transfer perpendicular to the CuO$_2$ bilayers indicates that the relevant spins are on Cu sites \cite{tran92} and not on O, where the doped holes are located \cite{nuck95}.

\begin{figure}[t]
\centerline{\includegraphics[width=6cm]{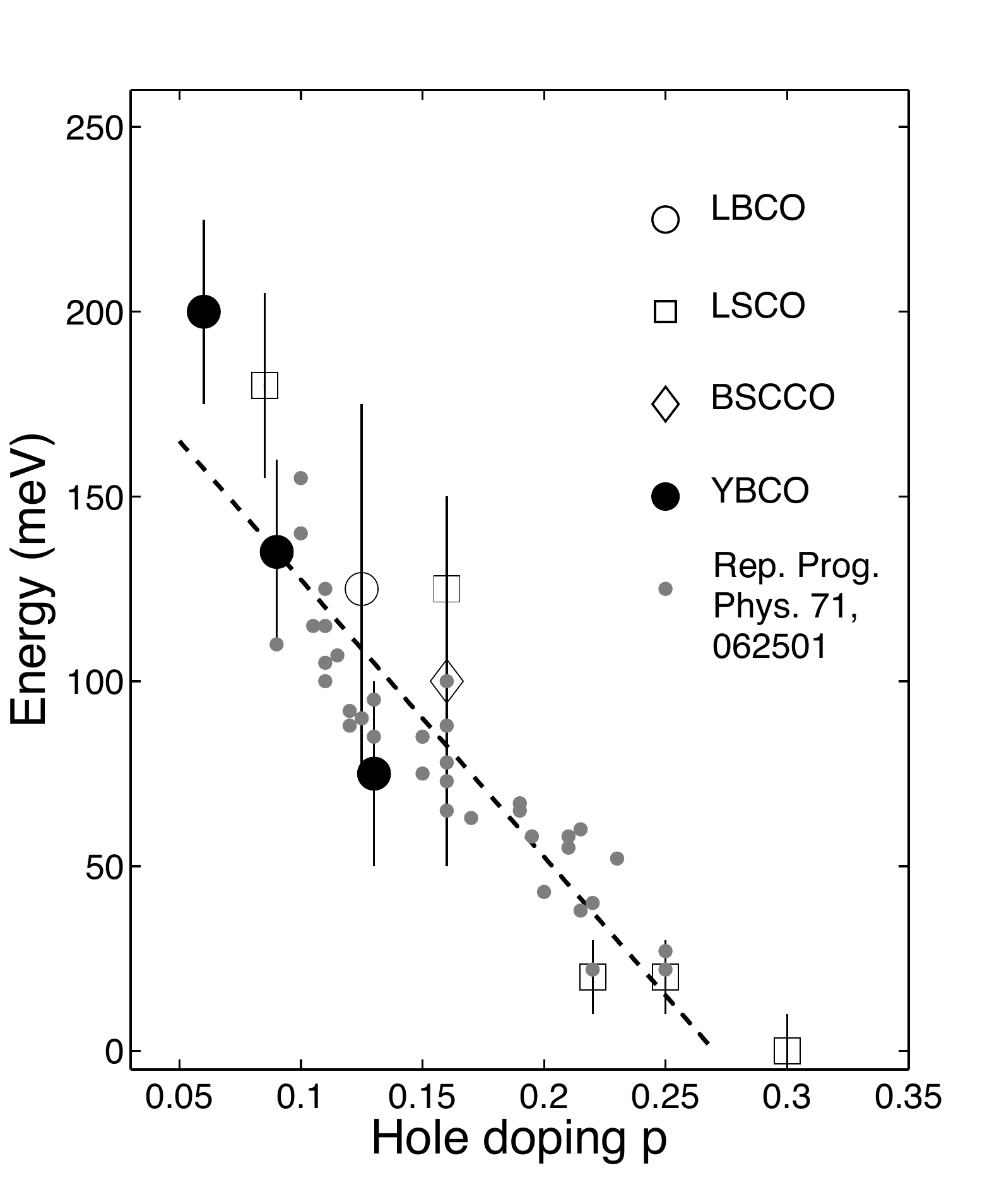}}
\caption{Large symbols: estimates of doping-dependent energy scale at which magnetic spectral weight falls below half that of the antiferrromagnetic state based on inelastic neutron scattering studies of various cuprates.  Small gray symbols: pseudogap energy from various electronic spectroscopies as summarized in \cite{hufn08}.  Reprinted figure with permission from Stock {\it et al.} \cite{stoc10}, \copyright\  2008 American Physical Society.}
\label{fg:pseudo}
\end{figure}

Another place to look for the impact of doped holes is at high excitation energies.  A number of groups have used time-of-flight measurements to characterize the spin fluctuations over a broad energy range in \ybco\ \cite{stoc07,pail04}, \lsco\ \cite{lips07,waki07b,lips09,chri04}, \lbco\ \cite{tran04,xu07}, and \bscco\ \cite{xu09}.    Analyzing the local susceptibility, $\chi''(\omega)$, Stock {\it et al.} \cite{stoc10} noticed that the strength of the magnetic response falls off dramatically and systematically above a doping-dependent energy scale.  To quantify this behavior, they evaluated the energy at which $\chi''(\omega)$ falls below half of that for an undoped antiferromagnet; their results are shown in Fig.~\ref{fg:pseudo}.  Remarkably, the identified energy scale corresponds very well with the pseudogap energy determined from electronic spectroscopies \cite{hufn08}.  The resulting picture is that spin fluctuations retain the strength of a correlated insulator below the pseudogap energy, where charge excitations are reduced, while magnetic weight is strongly depressed above the pseudogap, where charge excitations are stronger and better defined.

A recent resonant inelastic x-ray scattering (RIXS) study on \ybco\ and closely related compounds \cite{leta11} appears to conflict with these results.  First note that because of limitations on momentum transfer and energy resolution, the RIXS measurements are limited to ${\bf Q}=(h,0,0)$ with $h<0.45$ reciprocal lattice units and energies greater than 130 meV.  The excitations observed in an insulating sample are consistent with expectations for antiferromagnetic spin waves and appear consistent with neutron results.  This is reasonable, as the presence of antiferromagnetic order ensures that the dispersion must be the same about both ferromagnetic and antiferromagnetic zone centers.  Measurements on superconducting samples up to optimal doping, however, find excitations that soften no more than 10\%\ with doping, with negligible change in integrated intensity.  Such results are quite different from the neutron scattering results, especially those shown in Fig.~\ref{fg:pseudo}; they also conflict with 2-magnon Raman scattering results from Sugai {\it et al.} \cite{suga03}.  Of course, in the absence of antiferromagnetic order there is no constraint that the excitations, especially including the effects of damping, be the same in both the ferromagnetic and antiferromagnetic zones.   Now, the RIXS cross section includes both charge and spin excitations, and it is not yet firmly established how the charge and spin channels will interact in measurements of metallic samples.  Even assuming that a proper identification of magnetic excitations has been made, there is no physical justification for the assumption made in \cite{leta11} that the dispersion measured by RIXS near ${\bf Q}={\bf 0}$ is the same as that near the antiferromagnetic wave vector, where the neutron measurements have been done.   The neutron cross section is very well understood, and the indications of strong damping of large {\bf Q} magnetic excitations are clear.   As the large {\bf Q} excitations are the ones relevant to theory, the significance of the RIXS measurements is not clear.

\begin{figure}[t]
\centerline{\includegraphics[width=7cm]{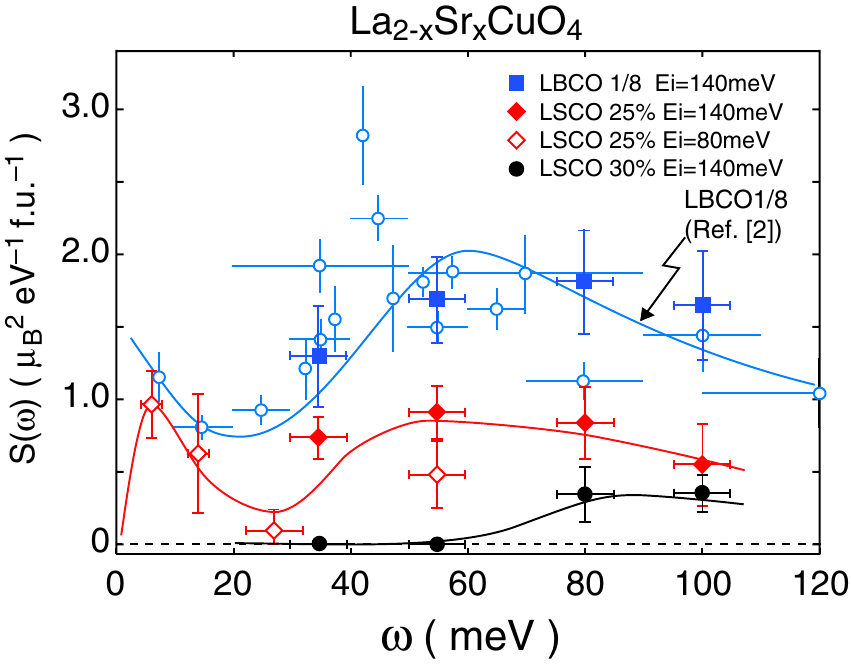}}
\caption{Comparison of magnetic intensity vs.\ energy in over-doped \lsco\ with $x=0.25$ (red diamonds) and $x=0.30$ (black circles) with \lbco\ with $x=0.125$ (blue squares).
Fom Wakimoto {\it et al.} \cite{waki07b}, \copyright\  2007 American Physical Society.}
\label{fg:overdoped}
\end{figure}

Implicit in Fig.~\ref{fg:pseudo} is the fact that AF excitations die away with over doping.  An explicit demonstration of this is given in Fig.~\ref{fg:overdoped} from \cite{waki07b}; related results for \lsco\ with $x=0.22$ have been reported by Lipscombe {\it et al.} \cite{lips07}.   Thus, as one dopes far away from the AF insulator state, the AF spin fluctuations disappear.  This occurs despite the fact that the Fermi surface becomes better nested with overdoping in \lsco\ \cite{yosh06}.

To summarize, experiments indicate that spin fluctuations in the cuprates evolve continuously from the AF insulator state.  Hole doping causes changes at low energy---incommensurate dispersion, spin gap, and resonance---and at high energy, where the spectral weight drops above the pseudogap energy; the AF fluctuations disappear with overdoping, where the superconductivity also goes away.  Model calculations in which the doped carriers are assumed to provide the magnetic response provide an appealingly direct connection to the superconductivity; however, they have a challenge to explain the doping dependence of the magnetic weight and the role of the superexchange energy $J$.  Alternatively, phase separation of the holes into charge stripes provides a natural way for locally-AF spin correlations among Cu moments to survive; the challenge here is to explain, from first principles, the transport properties and the superconductivity.  Empirically, the differences in the electronic properties of cuprates with stripe order and those without is extremely subtle \cite{vall06,he09,home12}.  Models of pairing \cite{scal12b} and superconductivity \cite{emer97} in striped systems have been proposed, but those models tend to involve {\it ad hoc} features.

Before leaving the cuprates, we should touch on the topic of ``intra-unit-cell'' magnetic order.  As reviewed by Bourges and Sidis \cite{bour11}, a proposal of loop-current order by Varma \cite{varm06} has motivated polarized neutron diffraction searches for a potential hidden order.  Varma's current loops are between a Cu atom and nearest-neighbor oxygens, so that the scattering should appear at Bragg peak positions; however, the size of the current loops implies a form factor that falls off rapidly with $Q$. In fact, the spatial extent of these current loops is similar to that of the strongly extended Wannier functions of magnetic electrons in cuprates, revealed by the covalent magnetic form factor \cite{ walt09}. Hence, the extent of the form factor of the loop contribution should be similar to that of the spin magnetic moment contribution, but the direction of the loop magnetic moment must be determined by its orbital origin.

Neutron studies have found evidence for a change in polarized neutron scattering for certain low-$Q$ peaks in several cuprate families below temperatures consistent with the onset of the pseudogap.  In addition, Li {\it et al.}\ \cite{li10} have reported a high-energy, dispersionless, apparently-magnetic mode in HgBa$_2$CuO$_6$, that has been interpreted as an Ising-like excitation of Varma's state \cite{he11}.  We note that Lederer and Kivelson \cite{lede12} have done an analysis of the loop current model and shown that, in combination with the neutron results, it implies finite magnetic fields at various lattice sites that should be detectable by nuclear magnetic resonance (NMR).  Definitive NMR tests have not yet been reported.

We are uncertain what to think of the measurements of intra-unit-cell magnetic order.  The measurements themselves are challenging to perform, as they require identifying a small magnetic signal on top of a substantial nuclear Bragg intensity. The observed polarization \cite{fauq06} of this additional contribution is at odds with the original model of current loops in the $ab$-plane \cite{varm06}. The situation would be easier to judge if there were other systems where this sort of magnetism has been observed.  (A recent experimental claim \cite{scag11} of orbital currents in antiferromagnetic CuO has been explained away in terms birefringence effects of the monoclinic lattice \cite{dima12}.)  In contrast, stripe order has been observed in a number of transition-metal oxides \cite{ulbr12b}, and, of course, there are examples of metallic antiferromagnets, such as Cr, where Fermi-surface nesting is important \cite{fawc88}.  If orbital currents are real, how do they interact with the spin response that has been clearly identified?

\section{Fe-based superconductors}

\subsection{Nature of the magnetism}

The crystal structure of iron pnictide and chalcogenide superconductors features a square-lattice arrangement of magnetic ions, similar to that in cuprates \cite{dela08,qiu08b,huan08a,li09b,lynn09,lums10r}. It consists of a continuous stacking of square-lattice layers of iron atoms, each sandwiched between the two half-density layers of bonding pnictogen or chalcogen anions. These anions, which tetrahedrally coordinate the Fe sites, occupy alternate checkerboard positions above and below the Fe layer, so that the resulting unit cell contains two formula units. The sparsely spaced layers are only weakly held together in this quasi-two-dimensional structure, which facilitates the reduced structural symmetry.

Unlike in the cuprates, where bridging oxygen anions connect the nearest-neighbor copper sites thus mediating the dominant nearest-neighbor superexchange interaction, bonding anions in the structure of iron-based superconductors connect the sites on the diagonal of the square plaquette. This suggests that the next-nearest neighbor, diagonal coupling might be the dominant superexchange interaction in the Heisenberg Hamiltonian of the local-spin picture for iron pnictides and chalcogenides \cite{si08,xuc08}. The resulting $J_1$--$J_2$ Heisenberg model exhibits frustration and has a complex phase diagram, including disordered spin-liquid and different N\'{e}el states \cite{chan90}. In addition, the nearest Fe-Fe distance is extraordinarily short, only $\approx 2.8$~\AA, suggesting that direct orbital overlaps could also be important, and so might be further-neighbor couplings. Extending the interactions to include coupling of second neighbors along the side of the square leads to a $J_1$--$J_2$--$J_3$ model on a square lattice \cite{more90}. Understanding different non-classical frustrated phases and the corresponding spin excitations in this model presents a major challenge. Thus, if we consider iron-based materials in terms of effective local-spin models, the situation is substantially more complex and less well understood than in the cuprates.

The most important distinction between the cuprates and the iron pnictides and chalcogenides, however, stems from the $3d^6$ electronic configuration of the Fe atom, compared to the $3d^9$ configuration of Cu \cite{tesa09}. Hence, while a single $d_{x^2-y^2}$ orbital is involved in describing the band structure of the cuprates, there are at least four active $d$-orbitals, in addition to the anion bonding $p$-orbitals, contributing to the unoccupied, or partially occupied band structure in the iron systems. In a strongly-correlated, localized-electron ionic picture, four $d$-orbitals would result in an $S = 2$ spin system, if Hund's rule is obeyed. Experiment, however, reveals maximum spins of $S=3/2$ in the most magnetic system, FeTe \cite{hu09b,zali12}, with smaller effective spins in ferropnictides \cite{lynn09,lums10r}. Half-integer spins immediately rule out the purely local-spin ionic picture with localized $d$-electrons, as only integer-spin states are allowed in this case. This, of course, agrees with the metallic conductivity in iron pnictides and chalcogenides observed in experiment.

It is perhaps worth recalling that the effective local spin is revealed in neutron scattering through the total scattering intensity. As given by Eq.~(\ref{eq:sum}), the integrated spectral weight corresponds to $S(S+1) = [\mu_{\rm eff}/(g \mu_B)]^2$, where $g$ is the Land\'e factor. This defines the fluctuating instantaneous effective moment, $\mu_{\rm eff}$, whereas the ordered static moment measured by Bragg diffraction in the magnetically ordered state, $\langle \mu \rangle = g \mu_B \langle S \rangle$, is always smaller. Here $\langle S \rangle$ is the ground-state value of the spin operator {\bf S}, and $g \mu_B \langle S \rangle < g \mu_B S < \mu_{\rm eff}$. In low-dimensional and/or frustrated systems, such as cuprates and iron pnictides and chalcogenides, $\langle S \rangle$ could be much smaller than the actual local spin $S$.

In the band theory description, even the number of electrons (holes) precludes iron-based materials from being Mott insulators and renders them semiconductors or semimetals with electron and hole pockets. Understanding the properties of these systems requires a multi-band description from the outset \cite{ande11}. Although many treatments of the itinerant-electron limit consider only two or three $t_{2g}$ derived bands formed by $d_{xz}$, $d_{yz}$ and $d_{xy}$ orbitals hybridized with the $p$-orbitals of the anion\cite{yin10,dagh10}, some researchers suggest that such treatment is insufficient and that more bands should be included \cite{yu11}. Theoretical results including all five $d$-orbital derived bands have been used to describe the neutron scattering data \cite{kane10,tham12,ewin11}. The weakness of the itinerant-electron description is that it only accounts for the on-site interactions by either applying perturbation theory \cite{erem10}, which fails when the interactions are substantial, or within the mean-field random phase approximation \cite{kane10}, which is not well controlled. Perhaps the most adequate way of treating the magnetic response of iron pnictides and chalcogenides theoretically, is by employing the dynamical mean field theory (DMFT), or a combination of DMFT with the density functional theory (DFT) \cite{haul09,yin11,aich10,park11b,liu12}. DMFT is designed to account for the on-site correlation, which in pnictides and chalcogenides is controlled to a large extent by the spin-dependent Hund's rule coupling \cite{haul09}.

\begin{figure*}[t]
\centerline{\includegraphics[width=0.7\linewidth]{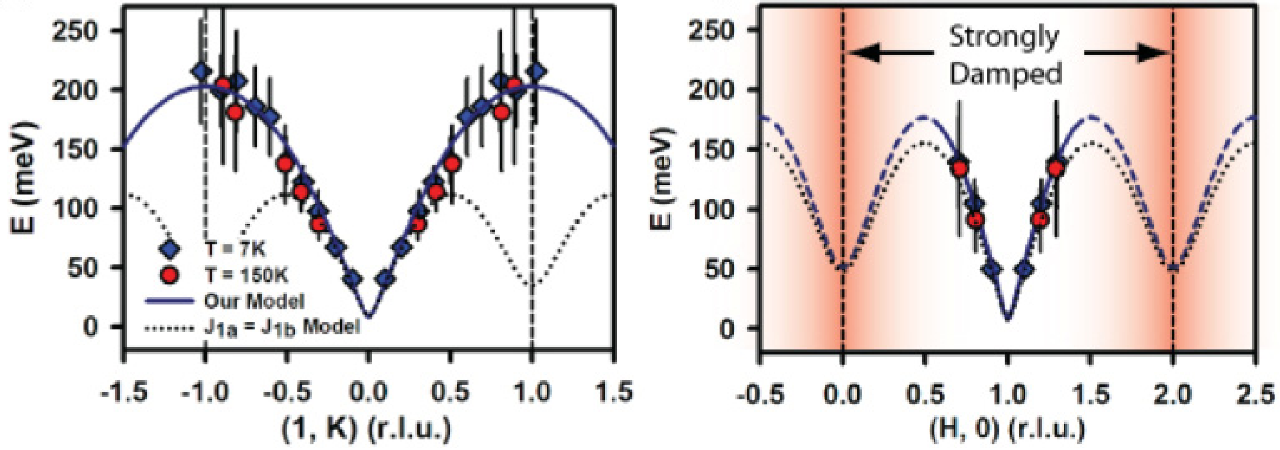}}
\caption{Magnetic excitations from BaFe$_2$As$_2$, modeled with a 
Heisenberg spin Hamiltonian by varying  exchange parameters $J_{1a}$, $J_{1b}$, 
and $J_2$, and applying an anisotropic damping. Reprinted figure with permission from Harriger {\it et al.}~\cite{harr11}, \copyright\  2011 American Physical Society.}
\label{fig:Dai2} 
\end{figure*}

To date, there have been numerous studies of the magnetic excitations in the antiferromagnetic members of various Fe-based superconductor families \cite{lynn09,lums10r,ewin11,liu12,dial09,wang11,harr11,zali11}.  In many papers, there is a focus on determining whether the excitations are better described in terms of the spin-wave spectrum derived from a spin-only Hamiltonian or of a weakly-coupling itinerant-calculation involving Fermi-surface nesting.  We argue that this is an ill-posed issue.  Calculations of a typical antiferromagnetically-ordered state within the local-spin density approximation indicate that there is spin polarization of states at high binding energies ($\sim 1 eV$) \cite{joha09}.  This method does not work in the paramagnetic state; however, attempts to calculate the phonon energies only find agreement with experiment when lattice constants are used that are consistent with those obtained in calculations with magnetic order \cite{yild09}.  (Spin-polarized ions are bigger than unpolarized ions.)  The electronic interactions, dominated by Hund's rule coupling, tend to be intermediate in strength \cite{yin10,yin11b}, so neither weak-coupling nor localized-moment approaches are accurate.

\begin{figure*}[t]
\centerline{\includegraphics[width=0.7\linewidth]{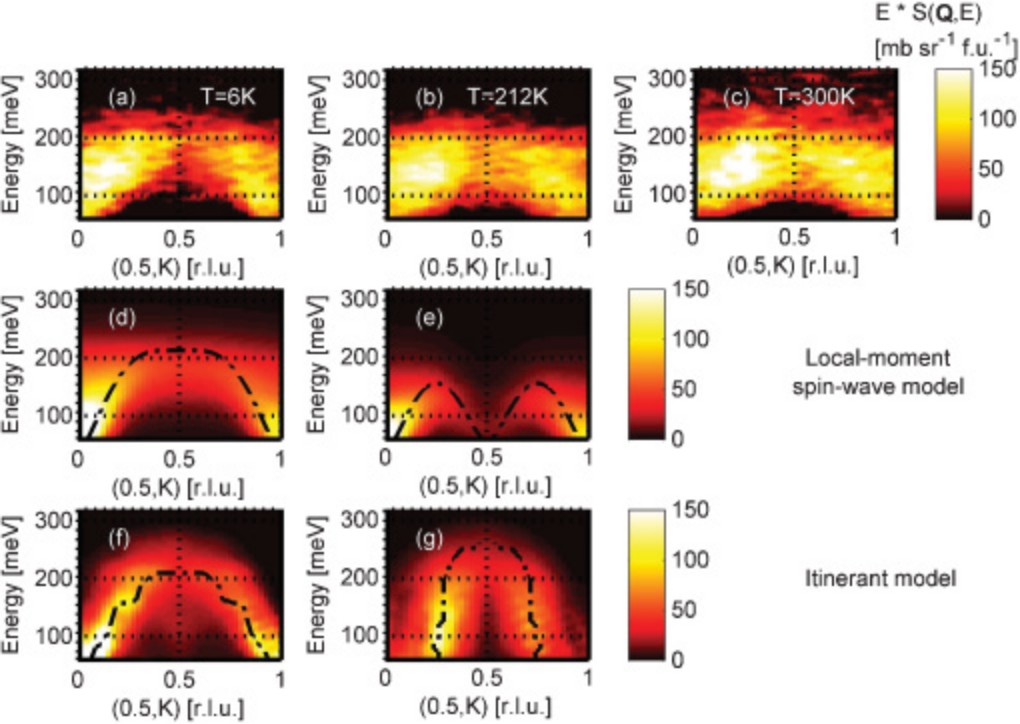}}
\caption{(a)-(c) Spectrum of magnetic excitations in SrFe$_2$As$_2$ at 6 K, 212 K and 300 K. (d) and (e) show calculations using the spin wave theory for local spins with anisotropic, $J_{1a} \neq J_{1b}$, and equal nearest-neighbor coupling, respectively. (f) and (g) are the results of the five band itinerant model calculation for the ordered and paramagnetic phase, respectively. Reprinted figure with permission from Ewings {\it et al.} \cite{ewin11}, \copyright\  2011 American Physical Society.}
\label{fg:Sr122spectrum}
\end{figure*}

Nevertheless, it is generally reasonable to use a spin-wave model based on a suitable spin-only Hamiltonian to parametrize measurements.  Figure~\ref{fig:Dai2} shows the spin excitation spectra measured by Harriger {\it et al.}\ \cite{harr11} in the AF phase of BaFe$_2$As$_2$.  As shown earlier for CaFe$_2$As$_2$ \cite{zhao09}, obtaining a spin-wave fit to the dispersion near the zone boundary along the [0,1] direction requires very different exchange parameters between nearest-neighbors along the spin direction and perpendicular to it.  Similar dispersions have been measured in the isostructural compound SrFe$_2$As$_2$ by Ewings {\it et al.}\ \cite{ewin11}, as shown in Fig.~\ref{fg:Sr122spectrum}.  In this figure, results are also shown for the paramagnetic state, which are quite similar to the ordered state.  Two different approaches to modeling the data are presented.  Clearly, if one uses a spin-wave model that is based on the symmetry of the magnetic state, with isotropic nearest-neighbor exchange in the paramagnetic phase, then the zone-boundary excitations cannot be properly described in the latter phase.  In contrast, a calculation for itinerant electrons within the random-phase approximation \cite{kane10} gets the zone-boundary excitations correctly, but has some discrepancy with the temperature dependence at lower energies.

Several groups have shown that the anisotropic nearest-neighbor exchange can occur due to breaking of the degeneracy between $d_{xz}$ and $d_{yz}$ orbitals \cite{krug09,lee09,chen09a}.  This orbital ordering \cite{lv09} is associated with the lowering of the lattice symmetry from tetragonal on cooling \cite{lynn09,lums10r}, and it has been verified by an angle-resolved photoemission study of Ba(Fe$_{1-x}$Co$_x$)$_2$As$_2$ \cite{yi11}. The structural transition and its relation to both the antiferromagnetic order and the superconductivity are of great interest and have been the subjects of intense study \cite{naka11,zhao08,li09b,chu09a,rotu10,nand10,ni10,kim11a,kim11b,mart11,bao09,rodr11,stoc11,li09a,mart10b,zali12}. Two general trends of the phase diagrams were established: (i) unless there is a first order magneto-structural transition, the lattice distortion usually occurs at a higher temperature ($T_s$) than the magnetic ordering ($T_N$), $T_s \gtrsim T_N$, and (ii) both $T_s$ and $T_N$ are reduced upon chemical substitution, so that both orders tend to disappear as the superconducting state develops.  The orbital ordering presumably occurs at $T_s$.  Experimental studies provide evidence for electronic anisotropy, often called nematic order, even in the tetragonal phase at temperatures not too far above $T_s$ \cite{chu12,yosh12}.  If the nematic order corresponds to correlated orbital occupancy, then the anisotropic magnetic exchange might still be applicable.  Alternatively, Wysocki {\it et al.}\ \cite{wyso11} have shown that similar effects can be modeled in an effective spin model by going beyond Heisenberg couplings (of the form $J{\bf S}_i\cdot{\bf S}_j$) to include a biquadratic term (of the form $-K({\bf S}_i\cdot{\bf S}_j)^2$).  

\begin{figure}[t]
\centerline{\includegraphics[width=1.\linewidth]{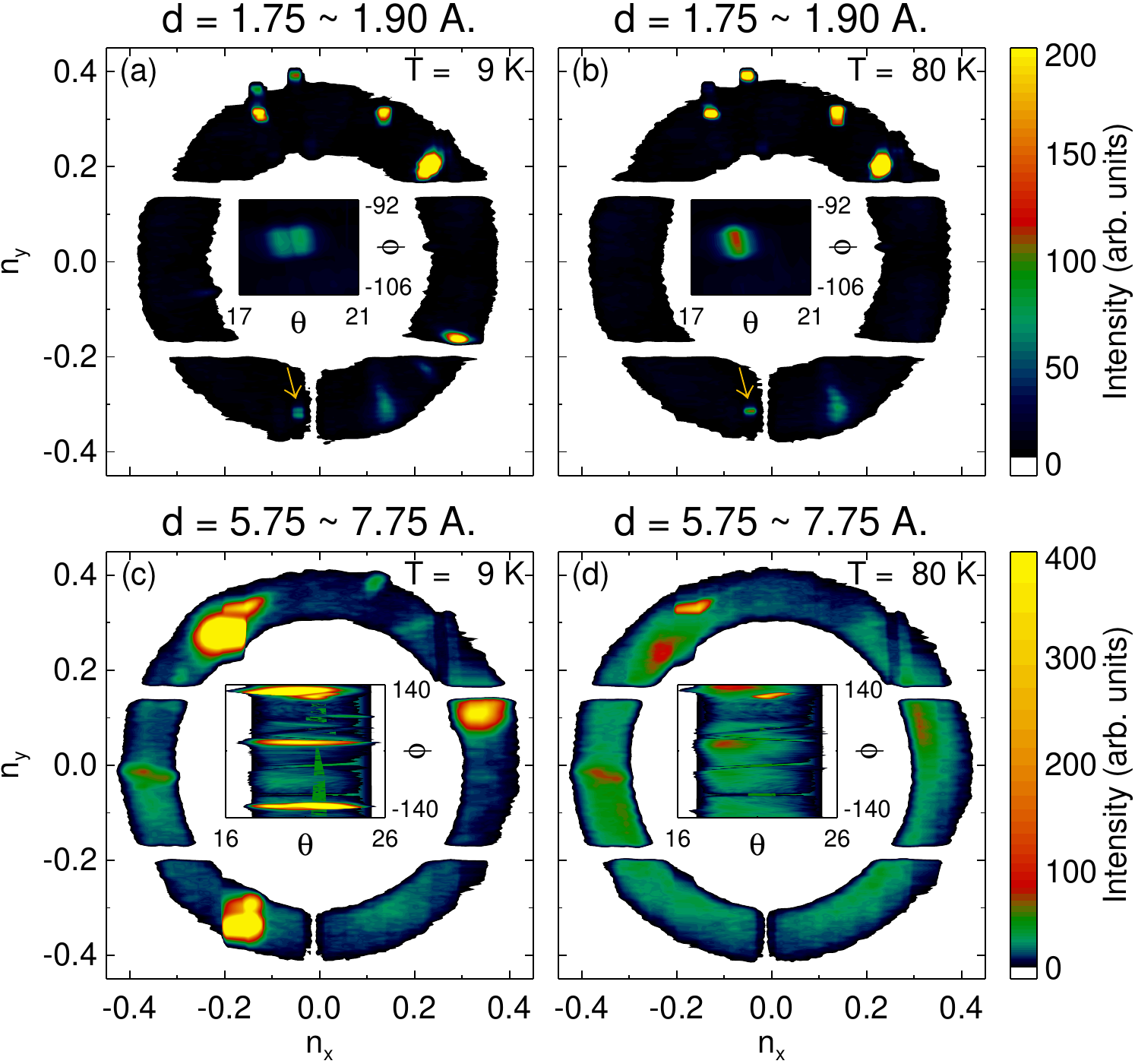}}
\caption{Maps of the scattered neutron intensity from Fe$_{1.1}$Te sample on the ARCS detector bank in the Laue mode as a function of the scattering direction at T = 9 K [(a) and (c)] and T = 80 K [(b) and (d)]. The top row shows structural scattering, while the bottom is magnetic scattering. From Zaliznyak {\it et al.} \cite{zali12}, \copyright\  2012 American Physical Society.}
\label{fg:FeTe_Laue}
\end{figure}

A very useful way of experimentally investigating the interplay of magnetic and structural transitions in a large single crystal sample used for neutron-spectroscopic study on the direct-geometry time-of-flight instrument is provided by deploying a quasi-Laue technique \cite{zali12}. In this mode, a quasi-white neutron beam with a broad band of incident neutron energies centered around some incident energy $E_i$ is selected by the pre-monochromating $T_0$ chopper. The crystal is aligned with respect to the incident beam direction so that Bragg reflections of interest appear on the detector. The detector signal is dominated by the elastic processes (diffraction), where the scattering angle is determined by the incident neutron energy (or wavelength, $\lambda_i$) and the $d$-spacing of the set of crystal planes involved in reflection, in accordance with Bragg's law, $\lambda_i = 2d \sin \theta$. Such a measurement is particularly well suited for studying the relative temperature evolution of structural and magnetic scattering, which are both present in the diffraction pattern at each $T$. Figure \ref{fg:FeTe_Laue} shows an example of such measurements performed with $E_i \approx 300$ meV  \cite{zali12}. 

\begin{figure}[t]
\centerline{\includegraphics[width=0.8\linewidth]{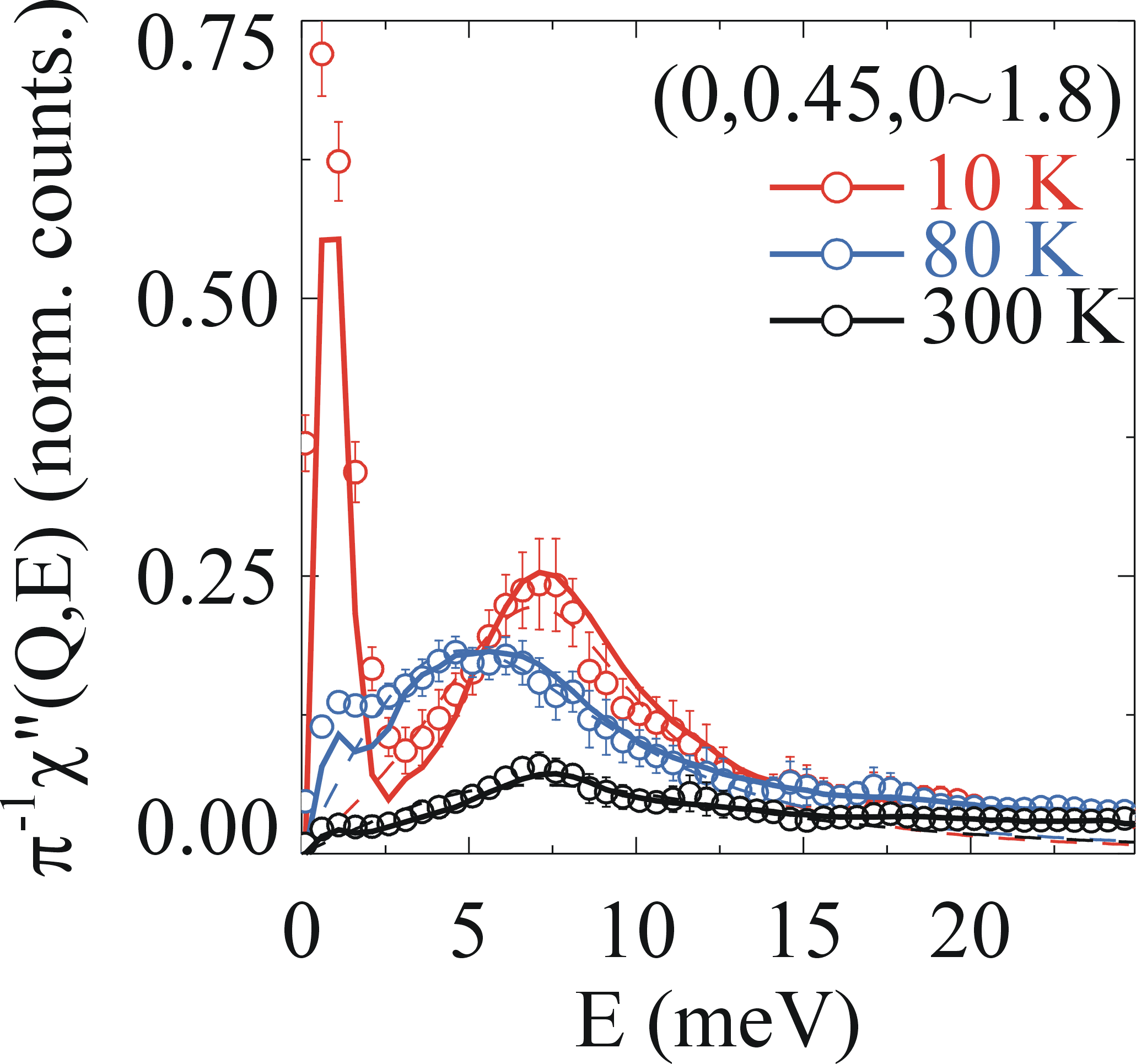}}
\caption{Energy dependence of the imaginary part of the dynamical magnetic susceptibility $\chi'' ({\bf Q}, E)$, near its maximum at $(h,k) \approx (0.5,0)$, in Fe$_{1.1}$Te at 10 K, 80 K and 300 K, from Zaliznyak {\it et al.} \cite{zali11}, \copyright\  2011 American Physical Society.}
\label{fg:FeTe_S(E)}
\end{figure}

The bipartite antiferromagnetic ordering observed in iron pnictide families is usually described as the spin-density wave (SDW) associated with the Fermi surface nesting of itinerant electrons \cite{mazi09,moon10,kuro08,dagh08,knol10}, which corresponds to the $(h,k) = (1/2,1/2)$ position in the $ab$-plane \cite{dela08,huan08a,qiu08b,li09b,lynn09,lums10r}. Such interpretation is based on the fact that the nesting wave vector matches that of the AF order, and on the small values of the observed ordered moment, which range from $\approx 0.9 \mu_B$ for NdFeAsO and BaFe$_2$As$_2$ \cite{qiu08b,huan08a}, to $\approx 0.4 \mu_B$ for LaOFeAs \cite{dela08}, and to only about $\approx 0.1 \mu_B$ in Na$_{1-\delta}$FeAs \cite{li09b}. Interestingly, the DFT band structure calculations typically overestimate the SDW order, predicting ordered moment consistent with nearly full spin polarization of an entire electronic band \cite{mazi09}.

On the other hand, the parent chalcogenide, Fe$_{1+y}$Te system, shows a ``bi-collinear'' magnetic ordering with the propagation vector $(h,k) = (1/2,0)$ in $P4/nmm$ reciprocal lattice units, which does not satisfy the nesting condition. The ground-state ordered moment, $\langle\mu \rangle \lesssim 2\mu_B$ \cite{bao09,rodr11,stoc11,li09a,mart10b}, although larger than in parent ferropnictides, is significantly smaller than the fully saturated value of $g \mu_B S$ (S = 3/2) corresponding to the fluctuating paramagnetic moment $\mu_{\rm eff} \approx 4\mu_B$ obtained from the Curie-Weiss behavior above 100 K \cite{hu09b,zali12}. In the local-spin picture, the low values of the AF ordered moment could arise from frustration and low-dimensionality of magnetic exchange interactions expected in these systems.

The magnetic form factor provides another important clue as to the nature of magnetism in these materials. It was recently studied by magnetic neutron diffraction in the antiferromagnetic SrFe$_2$As$_2$ \cite{lee10b,ratc10} and by the polarized neutron diffraction for both the AF BaFe$_2$As$_2$ \cite{brow10} and the superconducting Ba(Fe$_{1-x}$Co$_x$)$_2$As$_2$ \cite{prok11,lest11} compositions. The general features revealed by these studies are that (i) the magnetization density is only weakly anisotropic in wave vector, indicating that multiple $d$-orbitals of the iron atom contribute to the observed magnetic moments, and (ii) the $Q$-dependence of the form factor agrees with that for Fe atomic orbitals, indicating little hybridization of magnetic $d-$electrons with anions.

The full value of the fluctuating magnetic moment per Fe can be obtained by properly normalizing the spectral intensity of magnetic excitations measured in neutron experiment, and applying the sum rule for ${\cal S}(Q,E)$. Zaliznyak {\it et al.}\ \cite{zali11} carried out such a program for Fe$_{1.1}$Te and found that the low-energy part of magnetic fluctuations spectrum, for $E \lesssim 30$ meV, at 300 K already accounts for the local moment $\mu_{\rm eff} \approx 3.6\mu_B$. This is nearly consistent with $\mu_{\rm eff} \approx 3.87\mu_B$ corresponding to $S=3/2$ with $g \approx 2$, which is implicated in the Curie-Weiss behavior of the uniform magnetic susceptibility. The total neutron intensity, however, decreases roughly by a factor 2 upon cooling below $\approx 80$ K, corresponding to a change in the local spin value from $S = 3/2$ to $S = 1$. This could be viewed as an indication of the Kondo-like screening of the local spins by itinerant conduction electrons. The spectrum of $\chi''({\bf Q},E)$ in Fe$_{1.1}$Te near its maximum at ${\bf Q} = (h,k) \approx (0.5,0)$ at several temperatures is shown in Figure \ref{fg:FeTe_S(E)}.

\subsection{Magnetic correlations in the superconducting phase}

Magnetic order in the parent compound gradually diminishes with chemical 
doping, and superconductivity starts to emerge. In some systems the two phases
are well separated in the phase diagram~\cite{luet09,zhao08}, 
while in others they may coexist~\cite{chen09b, drew09,drew08}, either
on an atomic scale or through mesoscopic phase segregation. 
Eventually the AF phase is entirely 
suppressed and the system becomes a good superconductor at low temperature
below $T_c$.

\begin{figure*}[t]
\centerline{\includegraphics[width=\linewidth]{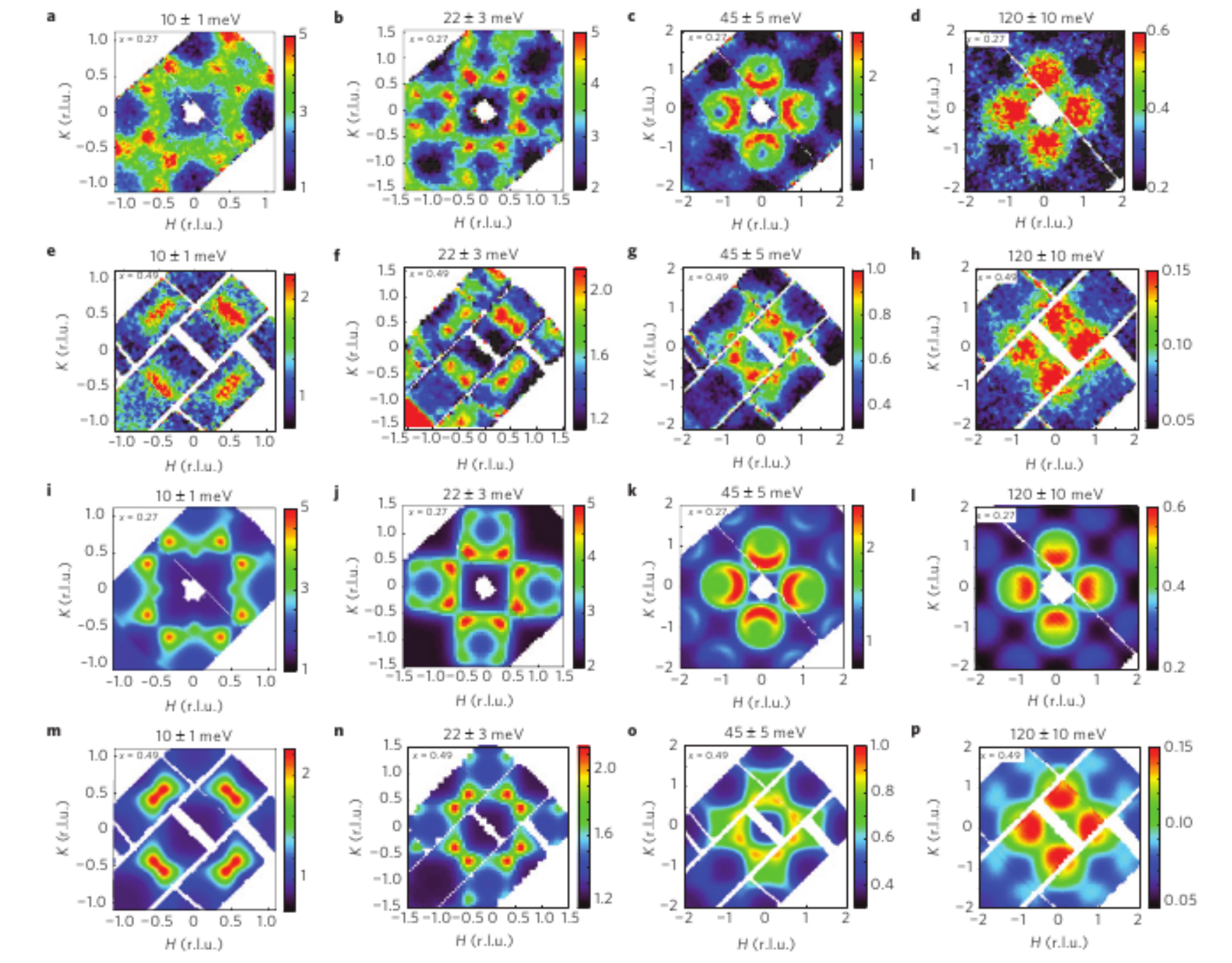}}
\caption{Constant-energy plots of the magnetic excitations in 
Fe$_{1+y}$Te$_{1-x}$Se$_x$ projected onto the $H$-$K$ plane. (a) - (d) are data taken from the $x=0.27$ non-superconducting
sample; (e) - (h) are data taken from the $x=0.49$ superconducting sample. 
(i) to (p) are model calculations based on the Sato-Maki function.  Reprinted by permission from Macmillan Publishers Ltd: Lumsden {\it et al.}~\cite{lums10}, \copyright\ 2010}
\label{fig:lumsden} 
\end{figure*}

For most of their energy band width, the spin fluctuations appear not to be strongly affected by superconductivity. An example is shown in Fig.~\ref{fig:lumsden}, where constant energy slices of magnetic scattering intensity from Fe$_{1+y}$Te$_{1-x}$Se$_x$  samples ($x=0.27$, non-superconducting; and $x=0.49$,superconducting) are plotted. The high energy excitations from 
the superconducting sample are qualitatively the same to those observed in the non-superconducting sample. A phenomenological model based on the Sato-Maki function \cite{sato74}, a form previously applied to studies of critical magnetic scattering in Cr \cite{noak90} and \lsco\ \cite{aepp97}, has been used to fit the data here, with parameters taken to be energy dependent.  

\begin{figure}[t]
\centerline{\includegraphics[width=1.\linewidth]{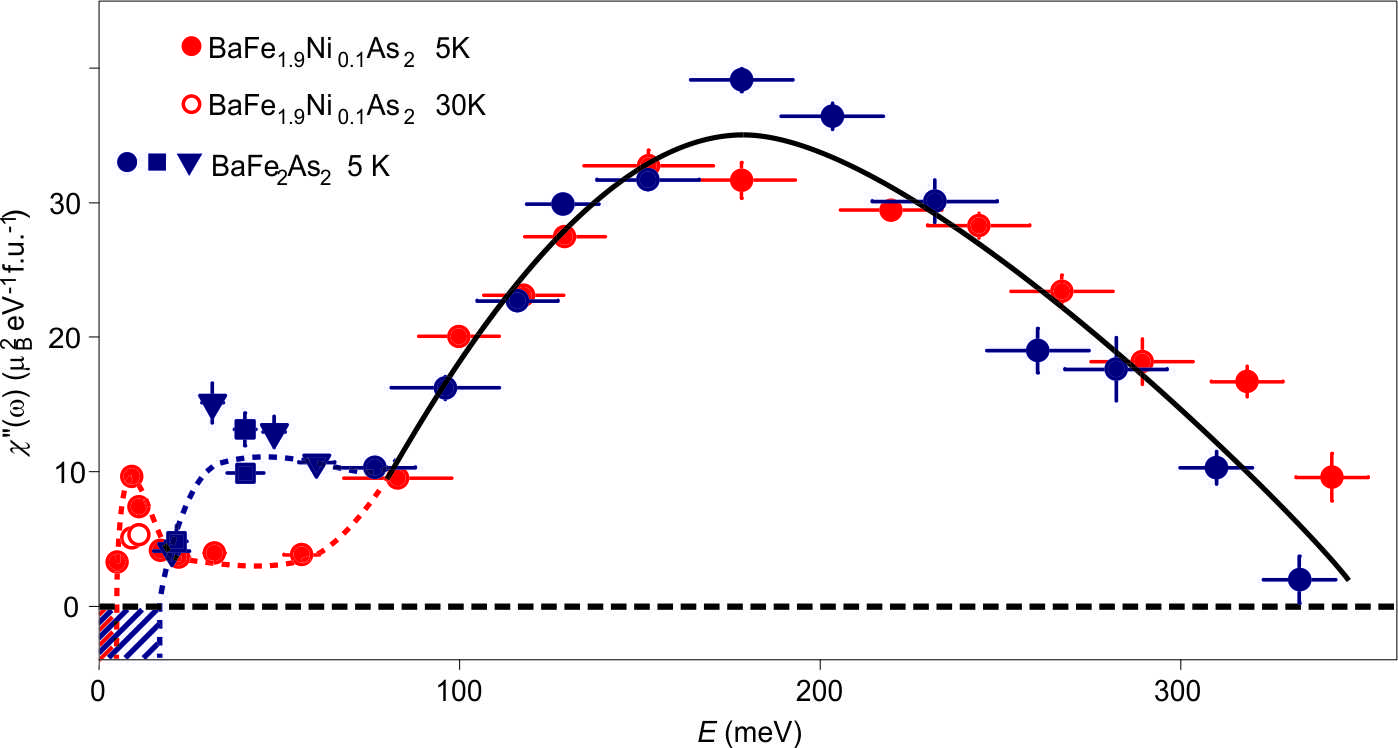}}
\caption{Energy dependence of the imaginary part of the dynamical local magnetic susceptibility $\chi'' (\omega)$ for BaFe$_2$As$_2$, and for BaFe$_{1.9}$Ni$_{0.1}$As$_2$ below (filled red circles) and above (open red circles) T$_c$.  Reprinted by permission from Macmillan Publishers Ltd:  Liu {\it et al.} \cite{liu12}, \copyright\ 2012.}
\label{fg:FePn_Chi(E)}
\end{figure}

A similar situation has been observed for high energy magnetic excitations in the 122 system.  Measurements on both the parent compound BaFe$_2$As$_2$~\cite{harr11} (or CaFe$_2$As$_2$~\cite{dial09,zhao09}) and superconducting Ba(Fe$_{1-x}$Co$_x$)$_2$As$_2$~\cite{ li10a,lest10} reveal similar magnetic dispersions.   The similarity between superconducting and parent compounds is especially clear in the comparison of {\bf Q}-integrated spectral weight between BaFe$_2$As$_2$ and BaFe$_{1.9}$Ni$_{0.1}$As$_2$ as obtained by Liu {\it et al.}\ \cite{liu12} and displayed in Fig.~\ref{fg:FePn_Chi(E)}.  As one can see, the spectral weight for spin excitations above 100~meV is virtually identical. The changes associated with doping and superconductivity are restricted to relatively small energies.  The integrated magnetic weight, however, does vary substantially among different Fe-based families.  In this case, the sum-rule weight corresponds to $\mu_{\rm eff} \approx 1.8\mu_B$ \cite{liu12,harr11}, which is much smaller than in iron telluride.  Within the experimental error this roughly corresponds to $S=1/2$, or one unpaired electron per Fe.

The temperature-dependent changes to the magnetic excitation spectra associated with the
development of superconductivity are similar to those in the
high-$T_c$ cuprates. In the superconducting phase, a ``spin resonance'' and
a  spin gap have been observed in various systems, such as the ``122'' compounds Ba$_{1-x}$K$_x$Fe$_2$As$_2$ \cite{chri08,cast11}, Ba(Fe$_{1-x}$Co$_x$)$_2$As$_2$ \cite{lums09,inos10}, Ba(Fe$_{1-x}$Ni$_{x}$)$_2$As$_2$ \cite{chi09,li09c}, and the ``11'' compound
Fe$_{1+\delta}$Te$_{1-x}$Se$_x$~\cite{qiu09, mook10}. 
In general the resonance energy $E_r$ scales with $T_c$ as 
$E_r \approx 4.3 \sim 5.3kT_c$~\cite{chri08,lums09,qiu09}; 
however this scaling appears to break down when pressure
is applied~\cite{mart12}.  The weak-coupling interpretation of the resonance and spin-gap in Fe-based systems is similar to that applied to the high-$T_c$ cuprates.  The existence of a resonance has been taken as evidence that the sign of the superconducting gap changes between Fermi surface pockets that are separated by the characteristic antiferromagnetic wave vector \cite{chub08,maie09}, although alternative perspectives have been advocated \cite{onar11}.

In both the 122 and 11 systems, the resonance always occurs around the in-plane wave vector 
(0.5,0.5) despite different ordered ground states in their parent compounds.  The excitations at higher energies appear to disperse only in the direction transverse to this wave vector.  It has been proposed that this unusual behavior might be due to the presence of orbital correlations \cite{lee10}; note that it has also been proposed \cite{kont10} that orbital fluctuations, rather than spin fluctuations, might mediate superconductivity.  In contrast, Castellan {\it et al.}\ \cite{cast11} have observed a longitudinal splitting of the spin resonance in Ba$_{1-x}$K$_x$Fe$_2$As$_2$ at high doping.  Theoretical RPA-type calculations taking account of imperfect Fermi surface nesting can reproduce this effect \cite{cast11}.

Both the intensity and energy of the resonance mode are found to be affected 
by temperature~\cite{inos10} and external magnetic field~\cite{wen10,zhao10}, 
demonstrating an intimate
correlation between the superconducting electron pairing and magnetic 
excitations in the Fe-based superconductors. Nevertheless, the microscopic
origin of the resonance mode is still not fully understood. Is it a singlet to 
triplet excitation arising from quasi-particle scattering, or simply a 
modification of existing magnetic excitations by the establishment of 
superconductivity? If the former is the case, a Zeeman splitting of the 
triplet mode is expected with the application of an external magnetic field.
In the Fe-based superconductors, because of the reduced $T_c$ 
compared to the cuprates, the resonance energy $E_r$ is also considerably lower,
making it easier to pursue this problem. A number of inelastic neutron 
scattering studies have been performed, aiming at understanding the change 
of the resonance mode under a magnetic field~\cite{wen10, zhao10,qiu09,bao10}, 
but so far there is  no conclusive evidence
on a possible Zeeman splitting and the question of the origin of 
the resonance remains open.

\begin{figure}[t]
\centerline{\includegraphics[width=\linewidth]{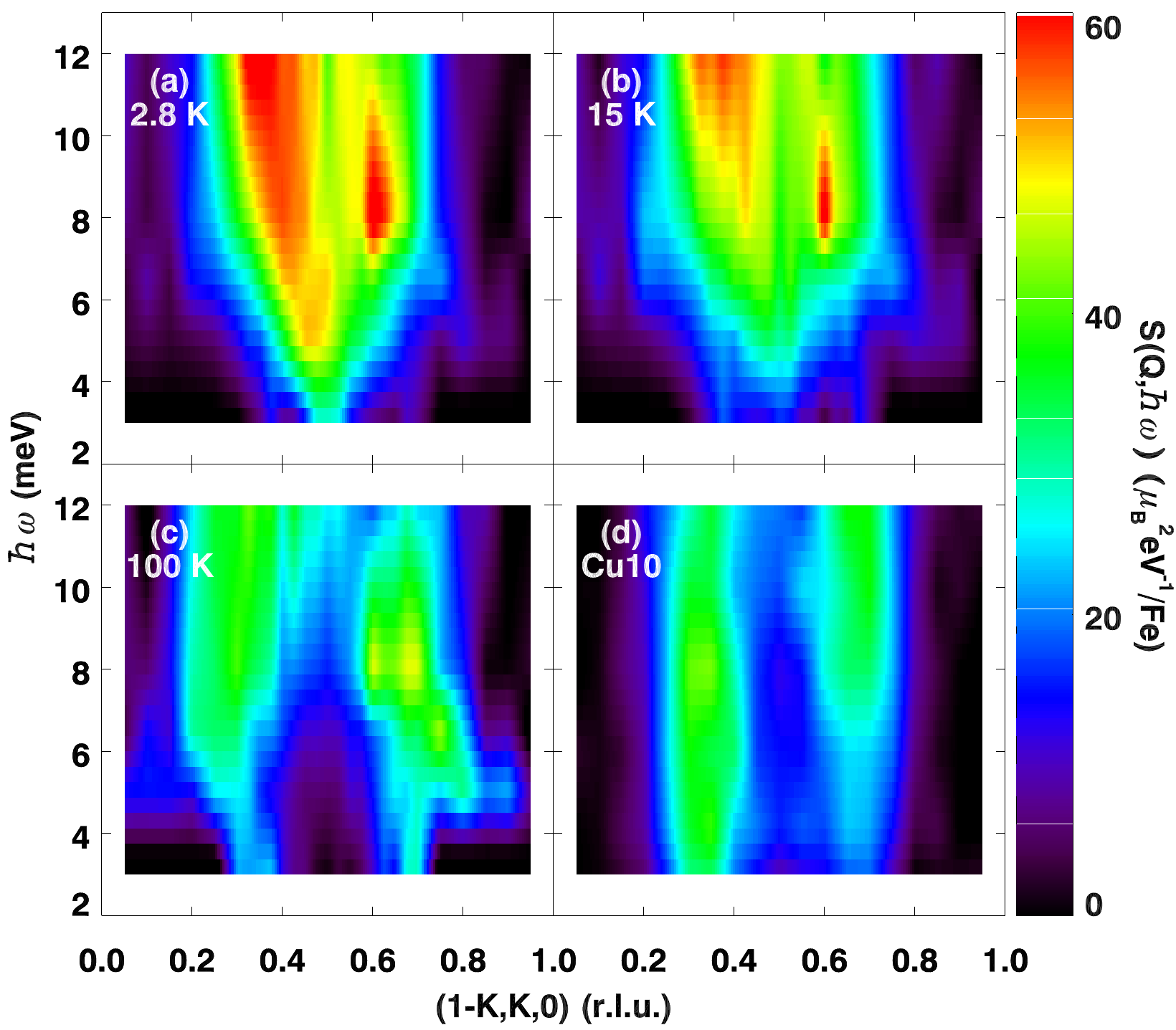}}
\caption{(a) to (c) Low energy magnetic excitations in a 
Fe$_{0.96}$Ni$_{0.04}$Te$_{0.5}$Se$_{0.5}$ superconducting ($T_C = 8$~K) sample.
(d) is the same data measured from a non-superconducting 
Fe$_{0.9}$Cu$_{0.1}$Te$_{0.5}$Se$_{0.5}$ sample, taken at 4~K; from 
Xu {\it et al.}~\cite{xu12a}.}
\label{fig:Xu1} 
\end{figure}

As mentioned previously, orbital order involving broken degeneracy of the $d_{xz}$ and $d_{yz}$ states has been predicted \cite{ krug09,lee09,chen09a}  and experimentally confirmed \cite{yi11} in the lowered-symmetry structural phase of iron pnictides and chalcogenides.  Sufficient chemical substitution leads to a restoration of tetragonal lattice symmetry, and hence orbital order should be suppressed; nevertheless, freezing of local orbital correlations 
may still occur~\cite{lee09,arha12}.  Although such effects are difficult to detect directly, they should have an impact on the magnetic response. Indeed a dramatic change in the dispersion
of low energy magnetic excitations has been observed in the superconducting 11 compound \cite{xu12a,tsyr12}, at a temperature of around 3$T_c$ (see Fig.~\ref{fig:Xu1}).  At about the same temperature scale, an abnormal in-plane expansion~\cite{xu12a, mart10b} is observed, which would be consistent with freezing of local orbital correlations. In addition, thermoelectric power~\cite{pall09}, 
and optical conductivity~\cite{home10,moon11} measurements
both indicate a change of electronic states near the Fermi level in this temperature range.  
Although the exact role of orbital correlations in the
magnetic and superconducting correlations are not fully
understood, this is a direction that deserves further attention.

Finally, we remind the reader that this is far from an exhaustive review of the field.   In particular, we note that there are some newer systems, such as (Li,Na)FeAs and A$_y$Fe$_{2-x}$Se$_2$ (A = K, Rb, Cs, Tl), that we have neglected for the sake of brevity.  For pointers to some of the neutron work on these very interesting materials, the reader may wish to consult the reviews \cite{dai12} and \cite{wen11}.

%%%%%%%%%%%%%%%%%%%%%%%%%

\section{Acknowledgements}

The authors are supported by the Office of Basic Energy Sciences, Division of Materials Science and Engineering, U.S. Department of Energy (DOE), under Contract No. DE-AC02-98CH10886. 

%\bibliographystyle{elsarticle-num}
%\bibliography{lno,theory,neutrons,fe_sc}

\begin{thebibliography}{100}
\expandafter\ifx\csname url\endcsname\relax
  \def\url#1{\texttt{#1}}\fi
\expandafter\ifx\csname urlprefix\endcsname\relax\def\urlprefix{URL }\fi
\expandafter\ifx\csname href\endcsname\relax
  \def\href#1#2{#2} \def\path#1{#1}\fi

\bibitem{mazi10}
I.~I. Mazin, %{Superconductivity gets an iron boost}, 
Nature 464 (2010)
  183--186.

\bibitem{lums10r}
M.~D. Lumsden, A.~D. Christianson, %{Magnetism in Fe-based superconductors}, 
J. Phys. Condens. Matter 22 (2010) 203203.

\bibitem{scal12a}
D.~J. Scalapino, %{A common thread: The pairing interaction for unconventiona  superconductors}, 
  Rev. Mod. Phys. 84 (2012) 1383--1417.

\bibitem{lee06}
P.~A. Lee, N.~Nagaosa, X.-G. Wen, %{Doping a Mott insulator: Physics of high-temperature superconductivity}, 
  Rev. Mod. Phys. 78 (2006) 17--85.

\bibitem{kive07}
S.~A. Kivelson, E.~Fradkin, %{How optimal inhomogeneity produces high temperature superconductivity}, 
  in: J.~R. Schrieffer, J.~S. Brooks (Eds.),
  Handbook of High-Temperature Superconductivity, Springer, New York, 2007, pp.
  570--596.

\bibitem{rita06}
K.~Lefmann, C.~Niedermayer, A.~Abrahamsen, C.~Bahl, N.~Christensen,
  H.~Jacobsen, T.~Larsen, P.~H{\"a}fliger, U.~Filges, H.~R{\o}nnow, %{Realizing the full potential of a RITA spectrometer}, 
  Physica B 385--386 (2006)
  1083--1085.

\bibitem{zali05}
I.~A. Zaliznyak, S.-H. Lee, {Magnetic Neutron Scattering}, in: Y.~Zhu (Ed.),
  Modern Techniques for Characterizing Magnetic Materials, Springer,
  Heidelberg, 2005.

\bibitem{macs08}
J.~A. Rodriguez, D.~M. Adler, P.~C. Brand, C.~Broholm, J.~C. Cook, C.~Brocker,
  R.~Hammond, Z.~Huang, P.~Hundertmark, J.~W. Lynn, N.~C. Maliszewskyj,
  J.~Moyer, J.~Orndorff, D.~Pierce, T.~D. Pike, G.~Scharfstein, S.~A. Smee,
  R.~Vilaseca, %{MACS---a new high intensity cold neutron spectrometer at NIST},
  Meas. Sci. Technol. 19 (2008) 034023.

\bibitem{lynn12}
J.~W. Lynn, Y.~Chen, S.~Chang, Y.~Zhao, S.~Chi, W.~R. II, B.~G. Ueland, R.~W.
  Erwin, J. Res. Nat. Inst. Stand. Technol. 117 (2012) 61--79.

\bibitem{maps94}
T.~G. Perring, A.~D. Taylor, R.~Osborn, D.~M. Paul, A.~T. Boothroyd, G.~Aeppli,
 % {MAPS: A chopper spectrometer to measure high energy magnetic excitations in single crystals},
   in: Proc. ICANS XII, RAL Report 94-025, 1994, pp. 1--60.

\bibitem{merlin06}
R.~Bewley, R.~Eccleston, K.~McEwen, S.~Hayden, M.~Dove, S.~Bennington,
  J.~Treadgold, R.~Coleman, %{MERLIN, a new high count rate spectrometer at ISIS}, 
  Physica B 385--386 (2006) 1029--1031.

\bibitem{arcs12}
D.~L. Abernathy, M.~B. Stone, M.~J. Loguillo, M.~S. Lucas, O.~Delaire, X.~Tang,
  J.~Y.~Y. Lin, B.~Fultz, %{Design and operation of the wide angular-range chopper spectrometer ARCS at the Spallation Neutron Source}, 
  Rev. Sci.
  Instrum. 83 (2012) 015114.

\bibitem{sequoia10}
G.~E. Granroth, A.~I. Kolesnikov, T.~E. Sherline, J.~P. Clancy, K.~A. Ross,
  J.~P.~C. Ruff, B.~D. Gaulin, S.~E. Nagler, %{SEQUOIA: A Newly Operating Chopper Spectrometer at the SNS}, 
  J. Phys. Conf. Ser. 251 (2010) 012058.

\bibitem{4seasons11}
R.~Kajimoto, M.~Nakamura, Y.~Inamura, F.~Mizuno, K.~Nakajima,
  S.~Ohira-Kawamura, T.~Yokoo, T.~Nakatani, R.~Maruyama, K.~Soyama, K.~Shibata,
  K.~Suzuya, S.~Sato, K.~Aizawa, M.~Arai, S.~Wakimoto, M.~Ishikado, S.~Shamoto,
  M.~Fujita, H.~Hiraka, K.~Ohoyama, K.~Yamada, C.-H. Lee, %{The Fermi Chopper Spectrometer 4SEASONS at J-PARC}, 
  J. Phys. Soc. Jpn. 80 (2011) SB025.

\bibitem{hyspec06}
S.~M. Shapiro, I.~A. Zaliznyak, L.~Passell, V.~J. Ghosh, W.~J. Leonhardt, M.~E.
  Hagen, %{HYSPEC: A crystal time-of-flight hybrid spectrometer for the spallation neutron source with polarization capabilities}, 
  Physica B 385--386
  (2006) 1107--1109.

\bibitem{regn03}
L.~P. Regnault, H.~M. {R{\o}nnow}, J.~E. Lorenzo, R.~Bellissent, F.~Tasset,
 % Inelastic neutron polarization analysis, 
  Physica B 335 (2003) 19--25.

\bibitem{fuji12a}
M.~Fujita, H.~Hiraka, M.~Matsuda, M.~Matsuura, J.~M. Tranquada, S.~Wakimoto,
  G.~Xu, K.~Yamada, %{Progress in Neutron Scattering Studies of Spin Excitations in High-$T_c$ Cuprates}, 
  J. Phys. Soc. Jpn. 81 (2012) 011007.

\bibitem{kast98}
M.~A. Kastner, R.~J. Birgeneau, G.~Shirane, Y.~Endoh, %Magnetic, transport, and optical properties of monolayer copper oxides, 
  Rev. Mod. Phys. 70 (1998) 897.

\bibitem{bour98}
P.~Bourges, in: J.~Bok, G.~Deutscher, D.~Pavuna, S.~A. Wolf (Eds.), The Gap
  Symmetry and Fluctuations in High Temperature Superconductors, Plenum, New
  York, 1998, p. 349.

\bibitem{maso01}
T.~E. Mason, in: J.~K.~A.~Gschneidner, L.~Eyring, M.~B. Maple (Eds.), Handbook
  on the Physics and Chemistry of Rare Earths, Vol.\ 31: High-Temperature
  Superconductors -- II, Elsevier, Amsterdam, 2001, pp. 281--314.

\bibitem{lynn01}
J.~W. Lynn, S.~Skanthakumar, in: J.~K.~A.~Gschneidner, L.~Eyring, M.~B. Maple
  (Eds.), Handbook on the Physics and Chemistry of Rare Earths, Vol. 31,
  Elsevier Science, Amsterdam, 2001, pp. 315--350.

\bibitem{tran07}
J.~M. Tranquada, %{Neutron Scattering Studies of Antiferromagnetic Correlations in Cuprates}, 
  in: J.~R. Schrieffer, J.~S. Brooks (Eds.), Handbook of
  High-Temperature Superconductivity, Springer, New York, 2007, pp. 257--298.

\bibitem{kive03}
S.~A. Kivelson, I.~P. Bindloss, E.~Fradkin, V.~Oganesyan, J.~M. Tranquada,
  A.~Kapitulnik, C.~Howald, %How to detect fluctuating stripes in the high-temperature superconductors, 
  Rev. Mod. Phys. 75 (2003) 1201.

\bibitem{deml04}
E.~Demler, W.~Hanke, S.-C. Zhang, %So(5) theory of antiferromagnetism and superconductivity, 
  Rev. Mod. Phys. 76 (2004) 909.

\bibitem{esch06}
M.~Eschrig, %{The effect of collective spin-1 excitations on electronic spectra in high-$T_{c}$ superconductors}, 
  Adv. Phys. 55 (2006) 47--183.

\bibitem{ogat08}
M.~Ogata, H.~Fukuyama, %{The $t$ -- $J$ model for the oxide high-$T_c$ superconductors}, 
  Rep. Prog. Phys. 71 (2008) 036501.

\bibitem{birg06}
R.~J. Birgeneau, C.~Stock, J.~M. Tranquada, K.~Yamada, %Magnetic neutron scattering in hole-doped cuprate superconductors, 
  J. Phys. Soc. Jpn. 75
  (2006) 111003.

\bibitem{lynn09}
J.~W. Lynn, P.~Dai, %Neutron studies of the iron-based family of high tc  magnetic superconductors, 
  Physica C 469 (2009) 469--476.

\bibitem{pagl10}
J.~Paglione, R.~L. Greene, %{High-temperature superconductivity in iron-based materials}, 
  Nat. Phys. 6 (2010) 645--658.

\bibitem{john10}
D.~C. Johnston, %{The puzzle of high temperature superconductivity in layered iron pnictides and chalcogenides}, 
  Adv. Phys. 59 (2010) 803--1061.

\bibitem{wen11}
J.~Wen, G.~Xu, G.~Gu, J.~M. Tranquada, R.~J. Birgeneau, %{Interplay between  magnetism and superconductivity in iron-chalcogenide superconductors: crystal growth and characterizations}, 
  Rep. Prog. Phys. 74 (2011) 124503.

\bibitem{dai12}
P.~Dai, J.~Hu, E.~Dagotto,
  %{{Magnetism and its microscopic origin in iron-based high-temperature superconductors}}, 
  Nat. Phys. 8 (2012)  709--718.

\bibitem{birg99}
R.~J. Birgeneau, M.~Greven, M.~A. Kastner, Y.~S. Lee, B.~O. Wells, Y.~Endoh,
  K.~Yamada, G.~Shirane, Phys. Rev. B 59 (1999) 13788.

\bibitem{keim92a}
B.~Keimer, A.~Aharony, A.~Auerbach, R.~J. Birgeneau, A.~Cassanho, Y.~Endoh,
  R.~W. Erwin, M.~A. Kastner, G.~Shirane, Phys. Rev. B 45 (1992) 7430.

\bibitem{yama87}
K.~Yamada, E.~Kudo, Y.~Endoh, Y.~Hidaka, M.~Oda, M.~Suzuki, T.~Murakami, Solid
  State Commun. 64 (1987) 753.

\bibitem{huck08}
M.~{H\"ucker}, G.~D. Gu, J.~M. Tranquada, %{Spin susceptibility of underdoped cuprate superconductors: Insights from a stripe-ordered crystal}, 
  Phys. Rev. B 78 (2008) 214507.

\bibitem{ande52}
P.~W. Anderson, %{An Approximate Quantum Theory of the Antiferromagnetic Ground State}, 
  Phys. Rev. 86 (1952) 694--701.

\bibitem{oguc60}
T.~Oguchi, %{Theory of Spin-Wave Interactions in Ferro- and Antiferromagnetism},
  Phys. Rev. 117 (1960) 117--123.

\bibitem{mano91}
E.~Manousakis, %{The spin-$\frac12$ Heisenberg antiferromagnet on a square  lattice and its application to the cuprous oxides}, 
  Rev. Mod. Phys. 63 (1991) 1--62.

\bibitem{head10}
N.~S. Headings, S.~M. Hayden, R.~Coldea, T.~G. Perring, %{Anomalous High-Energy Spin Excitations in the High-${T}_{c}$ Superconductor-Parent Antiferromagnet  ${\mathrm{La}}_{2}{\mathrm{CuO}}_{4}$}, 
  Phys. Rev. Lett. 105 (2010) 247001.

\bibitem{lore05}
J.~Lorenzana, G.~Seibold, R.~Coldea, Phys. Rev. B 72 (2005) 224511.

\bibitem{sand01}
A.~W. Sandvik, R.~R.~P. Singh, %{High-Energy Magnon Dispersion and Multimagnon  Continuum in the Two-Dimensional Heisenberg Antiferromagnet}, 
  Phys. Rev. Lett. 86 (2001) 528--531.

\bibitem{dall12}
B.~Dalla~Piazza, M.~Mourigal, M.~Guarise, H.~Berger, T.~Schmitt, K.~J. Zhou,
  M.~Grioni, H.~M. R\o{}nnow, %{Unified one-band Hubbard model for magnetic and electronic spectra of the parent compounds of cuprate superconductors}, 
  Phys. Rev. B 85 (2012) 100508.

\bibitem{zaan87}
J.~Zaanen, G.~A. Sawatzky, %{The electronic structure and superexchange  interactions in transition-metal compounds}, 
  Can. J. Phys. 65 (1987) 1262--1271.

\bibitem{roge89}
M.~Roger, J.~M. Delrieu, Phys. Rev. B 39 (1989) 2299.

\bibitem{sham93}
S.~Shamoto, M.~Sato, J.~M. Tranquada, B.~J. Sternlieb, G.~Shirane,
 % {Neutron-scattering study of antiferromagnetism in YBa$_2$Cu$_3$O$_{6.15}$},
  Phys. Rev. B 48 (1993) 13817--13825.

\bibitem{cold01}
R.~Coldea, S.~M. Hayden, G.~Aeppli, T.~G. Perring, C.~D. Frost, T.~E. Mason,
  S.-W. Cheong, Z.~Fisk, %{Spin Waves and Electronic Interactions in La$_2$CuO$_4$}, 
  Phys. Rev. Lett. 86 (2001) 5377--5380.

\bibitem{walt09}
A.~C. Walters, T.~G. Perring, J.-S. Caux, A.~T. Savici, G.~D. Gu, C.-C. Lee,
  W.~Ku, I.~A. Zaliznyak, %{Effect of covalent bonding on magnetism and the missing neutron intensity in copper oxide compounds}, 
  Nat. Phys. 5 (2009)  867--872.

\bibitem{wals01}
R.~E. Walstedt, S.-W. Cheong, %{Covalency in ${\mathrm{La}}_{2}{\mathrm{CuO}}_{4}:$ A study of ${}^{17}\mathrm{O}$  hyperfine couplings in the paramagnetic phase}, 
  Phys. Rev. B 64 (2001) 014404.

\bibitem{tran04}
J.~M. Tranquada, H.~Woo, T.~G. Perring, H.~Goka, G.~D. Gu, G.~Xu, M.~Fujita,
  K.~Yamada, %Quantum magnetic excitations from stripes in copper oxide  superconductors, 
  Nature 429 (2004) 534.

\bibitem{vign07}
B.~Vignolle, S.~M. Hayden, D.~F. McMorrow, H.~M. {R{\o}nnow}, B.~Lake, C.~D.
  Frost, T.~G. Perring, %{Two energy scales in the spin excitations of the high-temperature superconductor La$_{2-x}$Sr$_x$CuO$_4$}, 
  Nat. Phys. 3 (2007) 163.

\bibitem{hayd04}
S.~M. Hayden, H.~A. Mook, P.~Dai, T.~G. Perring, F.~Do\u{g}an, %The structure of the high-energy spin excitations in a high-transition-temperature superconductor, 
  Nature 429 (2004) 531.

\bibitem{stoc05}
C.~Stock, W.~J.~L. Buyers, R.~A. Cowley, P.~S. Clegg, R.~Coldea, C.~D. Frost,
  R.~Liang, D.~Peets, D.~Bonn, W.~N. Hardy, R.~J. Birgeneau, %{From incommensurate to dispersive spin-fluctuations: The high-energy inelastic spectrum in superconducting YBa$_2$Cu$_3$O$_{6.5}$}, 
  Phys. Rev. B 71 (2005) 024522.

\bibitem{stoc10}
C.~Stock, R.~A. Cowley, W.~J.~L. Buyers, C.~D. Frost, J.~W. Taylor, D.~Peets,
  R.~Liang, D.~Bonn, W.~N. Hardy, %{Effect of the pseudogap on suppressing high energy inelastic neutron scattering in superconducting YBa$_{2}$Cu$_{3}$O$_{6.5}$}, 
  Phys. Rev. B 82 (2010) 174505.

\bibitem{xu09}
G.~Xu, G.~D. Gu, M.~Hucker, B.~Fauque, T.~G. Perring, L.~P. Regnault, J.~M.
  Tranquada, %{Testing the itinerancy of spin dynamics in superconducting  Bi$_2$Sr$_2$CaCu$_2$O$_{8+\delta}$}, 
  Nat. Phys. 5 (2009) 642--646.

\bibitem{suga03}
S.~Sugai, H.~Suzuki, Y.~Takayanagi, T.~Hosokawa, N.~Hayamizu, Phys. Rev. B 68
  (2003) 184504.

\bibitem{bour97}
P.~Bourges, H.~F. Fong, L.~P. Regnault, J.~Bossy, C.~Vettier, D.~L. Milius,
  I.~A. Aksay, B.~Keimer, Phys. Rev. B 56 (1997) R11439.

\bibitem{bour99}
P.~Bourges, Y.~Sidis, H.~F. Fong, B.~Keimer, L.~P. Regnault, J.~Bossy, A.~S.
  Ivanov, D.~L. Milius, I.~A. Aksay, %{Spin dynamics in high-$T_c$  superconductors}, 
  AIP Conf. Proc. 483 (1999) 207.

\bibitem{sidi04}
Y.~Sidis, S.~Pailh{\`e}s, B.~Keimer, C.~Ulrich, L.~P. Regnault, Phys. Stat.
  Sol. (b) 241 (2004) 1204.

\bibitem{yu09}
G.~Yu, Y.~Li, E.~M. Motoyama, M.~Greven, %{A universal relationship between magnetic resonance and superconducting gap in unconventional superconductors}, 
  Nat. Phys. 5 (2009) 873--875.

\bibitem{viro90}
A.~Virosztek, J.~Ruvalds, {Nested-Fermi-liquid theory}, Phys. Rev. B 42 (1990)
  4064--4072.

\bibitem{fawc88}
E.~Fawcett, Rev. Mod. Phys. 60 (1988) 209.

\bibitem{tsue00}
C.~C. Tsuei, J.~R. Kirtley, %{Pairing symmetry in cuprate superconductors}, 
Rev. Mod. Phys. 72 (2000) 969--1016.

\bibitem{lu92}
J.~P. Lu, %{Neutron scattering as a probe for unconventional superconducting states}, 
  Phys. Rev. Lett. 68 (1992) 125--128.

\bibitem{hink10}
V.~Hinkov, C.~Lin, M.~Raichle, B.~Keimer, Y.~Sidis, P.~Bourges, S.~Pailh{\`e}s,
  A.~Ivanov, %{Superconductivity and electronic liquid-crystal states in  twin-free YBa$_2$Cu$_3$O$_{6+x}$ studied by neutron scattering}, 
  Eur. Phys. J. Special Topics 188 (2010) 113--129.

\bibitem{lake99}
B.~Lake, G.~Aeppli, T.~E. Mason, A.~Schr\"oder, D.~F. McMorrow, K.~Lefmann,
  M.~Isshiki, M.~Nohara, H.~Takagi, S.~M. Hayden, Nature 400 (1999) 43.

\bibitem{chri04}
N.~B. Christensen, D.~F. McMorrow, H.~M. R{\o}nnow, B.~Lake, S.~M. Hayden,
  G.~Aeppli, T.~G. Perring, M.~Mangkorntong, M.~Nohara, H.~Tagaki, Phys. Rev.
  Lett. 93 (2004) 147002.

\bibitem{zaan01}
J.~Zaanen, O.~Y. Osman, H.~V. Kruis, Z.~Nussinov, J.~Tworzyd{\l}o, %The  geometric order of stripes and luttinger liquids, 
  Phil. Mag. B 81 (2001) 1485--1531.

\bibitem{vojt09}
M.~Vojta, %{Lattice symmetry breaking in cuprate superconductors: Stripes, nematics, and superconductivity}, 
  Adv. Phys. 58 (2009) 699--820.

\bibitem{huck11}
M.~H\"ucker, M.~v.~Zimmermann, G.~D. Gu, Z.~J. Xu, J.~S. Wen, G.~Xu, H.~J.
  Kang, A.~Zheludev, J.~M. Tranquada, %{Stripe order in superconducting La$_{2-x}$Ba$_{x}$CuO$_{4}$ ($0.095\le{}x\le{}0.155$)}, 
  Phys. Rev. B 83 (2011) 104506.

\bibitem{li07}
Q.~Li, M.~{H\"ucker}, G.~D. Gu, A.~M. Tsvelik, J.~M. Tranquada,
  %{Two-Dimensional Superconducting Fluctuations in Stripe-Ordered  La$_{1.875}$Ba$_{0.125}$CuO$_4$}, 
  Phys. Rev. Lett. 99 (2007) 067001.

\bibitem{tran08}
J.~M. Tranquada, G.~D. Gu, M.~H{\"u}cker, Q.~Jie, H.-J. Kang, R.~Klingeler,
  Q.~Li, N.~Tristan, J.~S. Wen, G.~Y. Xu, Z.~J. Xu, J.~Zhou, M.~v.~Zimmermann,
 % {Evidence for unusual superconducting correlations coexisting with stripe order in La$_{1.875}$Ba$_{0.125}$CuO$_4$}, 
  Phys. Rev. B 78 (2008) 174529.

\bibitem{steg12}
Z.~Stegen, S.~J. Han, J.~Wu, G.~D. Gu, Q.~Li, J.~H. Park, G.~S. Boebinger,
  J.~M. Tranquada, {Evolution of superconducting correlations within
  magnetic-field-decoupled CuO$_2$ layers of La$_{1.905}$Ba$_{0.095}$CuO$_4$}.
\newblock \href {http://arxiv.org/abs/arXiv:1207.0528}
  {\path{arXiv:1207.0528}}.

\bibitem{berg09b}
E.~Berg, E.~Fradkin, S.~A. Kivelson, J.~M. Tranquada, %{Striped superconductors: How the cuprates intertwine spin, charge and superconducting orders}, 
  New J. Phys. 11 (2009) 115004.

\bibitem{fuji04}
M.~Fujita, H.~Goka, K.~Yamada, J.~M. Tranquada, L.~P. Regnault, %{Stripe order, depinning, and fluctuations in La$_{1.875}$Ba$_{0.125}$CuO$_4$ and La$_{1.875}$Ba$_{0.075}$Sr$_{0.050}$CuO$_4$}, 
  Phys. Rev. B 70 (2004) 104517.

\bibitem{xu07}
G.~Y. Xu, J.~M. Tranquada, T.~G. Perring, G.~D. Gu, M.~Fujita, K.~Yamada,
  %{High-energy magnetic excitations from dynamic stripes in La$_{1.875}$Ba$_{0.125}$CuO$_4$}, 
  Phys. Rev. B 76 (2007) 014508.

\bibitem{aepp97}
G.~Aeppli, T.~E. Mason, S.~M. Hayden, H.~A. Mook, J.~Kulda, Science 278 (1997)
  1432.

\bibitem{tran92}
J.~M. Tranquada, P.~M. Gehring, G.~Shirane, S.~Shamoto, M.~Sato,
 % {Neutron-scattering study of the dynamical spin susceptibility in YBa$_2$Cu$_3$O$_{6.6}$}, 
  Phys. Rev. B 46 (1992) 5561--5575.

\bibitem{nuck95}
N.~N\"ucker, E.~Pellegrin, P.~Schweiss, J.~Fink, S.~L. Molodtsov, C.~T.
  Simmons, G.~Kaindl, W.~Frentrup, A.~Erb, G.~M\"uller-Vogt, %{Site-specific and doping-dependent electronic structure of ${\mathrm{YBa}}_{2}$${\mathrm{Cu}}_{3}$${\mathrm{O}}_{\mathit{x}}$ probed by O 1 \textit{s} and Cu 2 \textit{p} x-ray-absorption spectroscopy}, 
  Phys. Rev. B 51 (1995) 8529--8542.

\bibitem{stoc07}
C.~Stock, R.~A. Cowley, W.~J.~L. Buyers, R.~Coldea, C.~Broholm, C.~D. Frost,
  R.~J. Birgeneau, R.~Liang, D.~Bonn, W.~N. Hardy, %{Evidence for decay of spin  waves above the pseudogap of underdoped YBa$_{2}$Cu$_{3}$O$_{6.35}$}, 
  Phys. Rev. B 75 (2007) 172510.

\bibitem{pail04}
S.~Pailh\`es, Y.~Sidis, P.~Bourges, V.~Hinkov, A.~Ivanov, C.~Ulrich, L.~P.
  Regnault, B.~Keimer, Phys. Rev. Lett. 93 (2004) 167001.

\bibitem{lips07}
O.~J. Lipscombe, S.~M. Hayden, B.~Vignolle, D.~F. McMorrow, T.~G. Perring,
 % {Persistence of High-Frequency Spin Fluctuations in Overdoped Superconducting La$_{2-x}$Sr$_{x}$CuO$_4$ ($x = 0.22$)}, 
  Phys. Rev. Lett. 99 (2007) 067002.

\bibitem{waki07b}
S.~Wakimoto, K.~Yamada, J.~M. Tranquada, C.~D. Frost, R.~J. Birgeneau,
  H.~Zhang, %{Disappearance of Antiferromagnetic Spin Excitations in Overdoped La$_{2-x}$Sr$_x$CuO$_4$}, 
  Phys. Rev. Lett. 98 (2007) 247003.

\bibitem{lips09}
O.~J. Lipscombe, B.~Vignolle, T.~G. Perring, C.~D. Frost, S.~M. Hayden,
 % {Emergence of Coherent Magnetic Excitations in the High Temperature Underdoped La$_{2-x}$Sr$_{x}$CuO4$_{4}$ Superconductor at Low Temperatures},
  Phys. Rev. Lett. 102 (2009) 167002.

\bibitem{hufn08}
S.~{H\"{u}fner}, M.~A. Hossain, A.~Damascelli, G.~A. Sawatzky, %Two gaps make a high-temperature superconductor?, 
  Rep. Prog. Phys. 71 (2008) 062501.

\bibitem{leta11}
M.~Le~Tacon, G.~Ghiringhelli, J.~Chaloupka, M.~M. Sala, V.~Hinkov, M.~W.
  Haverkort, M.~Minola, M.~Bakr, K.~J. Zhou, S.~Blanco-Canosa, C.~Monney, Y.~T.
  Song, G.~L. Sun, C.~T. Lin, G.~M. De~Luca, M.~Salluzzo, G.~Khaliullin,
  T.~Schmitt, L.~Braicovich, B.~Keimer, %{Intense paramagnon excitations in a  large family of high-temperature superconductors}, 
  Nat. Phys. 7 (2011) 725--730.

\bibitem{yosh06}
T.~Yoshida, X.~J. Zhou, K.~Tanaka, W.~L. Yang, Z.~Hussain, Z.-X. Shen,
  A.~Fujimori, S.~Sahrakorpi, M.~Lindroos, R.~S. Markiewicz, A.~Bansil,
  S.~Komiya, Y.~Ando, H.~Eisaki, T.~Kakeshita, S.~Uchida, %{Systematic doping evolution of the underlying Fermi surface of La$_{2-x}$Sr$_{x}$CuO$_{4}$},
  Phys. Rev. B 74 (2006) 224510.

\bibitem{vall06}
T.~Valla, A.~V. Federov, J.~Lee, J.~C. Davis, G.~D. Gu, %{The Ground State of the Pseudogap in Cuprate Superconductors}, 
  Science 314 (2006) 1914.

\bibitem{he09}
R.-H. He, K.~Tanaka, S.-K. Mo, T.~Sasagawa, M.~Fujita, T.~Adachi, N.~Mannella,
  K.~Yamada, Y.~Koike, Z.~Hussain, Z.-X. Shen, %{Energy gaps in the failed high-$T_c$ superconductor La$_{1.875}$Ba$_{0.125}$CuO$_4$}, 
  Nat. Phys. 5 (2009) 119--123.

\bibitem{home12}
C.~C. Homes, M.~H\"ucker, Q.~Li, Z.~J. Xu, J.~S. Wen, G.~D. Gu, J.~M.
  Tranquada, %{Determination of the optical properties of  La$_{2-x}$Ba$_{x}$CuO$_{4}$ for several dopings, including the anomalous  $x=\frac{1}{8}$ phase}, 
  Phys. Rev. B 85 (2012) 134510.

\bibitem{scal12b}
D.~Scalapino, S.~White, %{Stripe structures in the t-t'-J model}, 
Physica C 481 (2012) 146--152.

\bibitem{emer97}
V.~J. Emery, S.~A. Kivelson, O.~Zachar, Phys. Rev. B 56 (1997) 6120.

\bibitem{bour11}
P.~Bourges, Y.~Sidis, %{Novel magnetic order in the pseudogap state of high- copper oxides superconductors}, 
  C. R. Phys. 12 (2011) 461--479.

\bibitem{varm06}
C.~M. Varma, %{Theory of the pseudogap state of the cuprates}, 
Phys. Rev. B 73 (2006) 155113.

\bibitem{li10}
L.~Li, Y.~Wang, S.~Komiya, S.~Ono, Y.~Ando, G.~D. Gu, N.~P. Ong, %{Diamagnetism and Cooper pairing above $T_c$ in cuprates}, 
  Phys. Rev. B 81 (2010) 054510.

\bibitem{he11}
Y.~He, C.~M. Varma, %{Collective Modes in the Loop Ordered Phase of Cuprate Superconductors}, 
  Phys. Rev. Lett. 106 (2011) 147001.

\bibitem{lede12}
S.~Lederer, S.~A. Kivelson, %{Observable NMR signal from circulating current order in YBCO}, 
  Phys. Rev. B 85 (2012) 155130.

\bibitem{fauq06}
B.~Fauqu\'e, Y.~Sidis, V.~Hinkov, S.~Pailh\`es, C.~T. Lin, X.~Chaud,
  P.~Bourges, %{Magnetic Order in the Pseudogap Phase of High-$T_{c}$ Superconductors}, 
  Phys. Rev. Lett. 96 (2006) 197001.

\bibitem{scag11}
V.~Scagnoli, U.~Staub, Y.~Bodenthin, R.~A. de~Souza, M.~Garc\'{\i}a-Fern\'andez, M.~Garganourakis, A.~T. Boothroyd, D.~Prabhakaran, S.~W.
  Lovesey, %{Observation of Orbital Currents in CuO}, 
  Science 332 (2011) 696--698.

\bibitem{dima12}
S.~{Di Matteo}, M.~R. Norman, %{Orbital currents, anapoles, and magnetic quadrupoles in CuO}, 
  Phys. Rev. B 85 (2012) 235143.

\bibitem{ulbr12b}
H.~Ulbrich, M.~Braden, %{Neutron scattering studies on stripe phases in  non-cuprate materials}, 
  Physica C 481 (2012) 31--45.

\bibitem{dela08}
C.~de~la Cruz, Q.~Huang, J.~W. Lynn, J.~Li, W.~R. Ii, J.~L. Zarestky, H.~A.
  Mook, G.~F. Chen, J.~L. Luo, N.~L. Wang, P.~Dai, %Magnetic order close to superconductivity in the iron-based layered lao1-xfxfeas systems, 
  Nature 453 (2008) 899--902.

\bibitem{qiu08b}
Y.~Qiu, W.~Bao, Q.~Huang, T.~Yildirim, J.~M. Simmons, M.~A. Green, J.~W. Lynn,
  Y.~C. Gasparovic, J.~Li, T.~Wu, G.~Wu, X.~H. Chen, %{Crystal Structure and Antiferromagnetic Order in ${\mathrm{NdFeAsO}}_{1-x}{\mathbf{F}}_{x}$ ($x=0.0$ and 0.2) Superconducting Compounds from Neutron Diffraction Measurements}, 
  Phys. Rev. Lett. 101 (2008) 257002.

\bibitem{huan08a}
Q.~Huang, Y.~Qiu, W.~Bao, M.~A. Green, J.~W. Lynn, Y.~C. Gasparovic, T.~Wu,
  G.~Wu, X.~H. Chen, %{Neutron-Diffraction Measurements of Magnetic Order and a Structural Transition in the Parent ${\mathrm{BaFe}}_{2}{\mathrm{As}}_{2}$ Compound of FeAs-Based High-Temperature Superconductors}, 
  Phys. Rev. Lett. 101 (2008) 257003.

\bibitem{li09b}
S.~Li, C.~de~la Cruz, Q.~Huang, G.~F. Chen, T.-L. Xia, J.~L. Luo, N.~L. Wang,
  P.~Dai, %{Structural and magnetic phase transitions in Na$_{1-\delta}$FeAs},
  Phys. Rev. B 80 (2009) 020504(R).

\bibitem{si08}
Q.~Si, E.~Abrahams, %{Strong Correlations and Magnetic Frustration in the High ${T}_{c}$ Iron Pnictides}, 
  Phys. Rev. Lett. 101 (2008) 076401.

\bibitem{xuc08}
C.~Xu, M.~M\"uller, S.~Sachdev, %{Ising and spin orders in the iron-based superconductors}, 
  Phys. Rev. B 78 (2008) 020501.

\bibitem{chan90}
P.~Chandra, P.~Coleman, A.~I. Larkin, %{Ising transition in frustrated Heisenberg models}, 
  Phys. Rev. Lett. 64 (1990) 88--91.

\bibitem{more90}
A.~Moreo, E.~Dagotto, T.~Jolicoeur, J.~Riera, %{Incommensurate correlations in the \textit{t} - \textit{J} and frustrated spin-1/2 Heisenberg models}, 
  Phys. Rev. B 42 (1990) 6283--6293.

\bibitem{tesa09}
Z.~Tesanovic, %Are iron pnictides new cuprates?, 
Physics 2 (2009) 60.

\bibitem{hu09b}
R.~Hu, E.~S. Bozin, J.~B. Warren, C.~Petrovic, %{Superconductivity, magnetism,  and stoichiometry of single crystals of Fe$_{1+y}$(Te$_{1-x}$S$_x$)$_z$},
  Phys. Rev. B 80 (2009) 214514.

\bibitem{zali12}
I.~A. Zaliznyak, Z.~J. Xu, J.~S. Wen, J.~M. Tranquada, G.~D. Gu, V.~Solovyov,
  V.~N. Glazkov, A.~I. Zheludev, V.~O. Garlea, M.~B. Stone, %{Continuous magnetic and structural phase transitions in Fe${}_{1+y}$Te}, 
  Phys. Rev. B 85 (2012) 085105.

\bibitem{ande11}
O.~Andersen, L.~Boeri, %{On the multi-orbital band structure and itinerant magnetism of iron-based superconductors}, 
  Ann. Phys. (Leipzig) 523 (2011) 8--50.

\bibitem{yin10}
W.-G. Yin, C.-C. Lee, W.~Ku, %{Unified Picture for Magnetic Correlations in Iron-Based Superconductors}, 
  Phys. Rev. Lett. 105 (2010) 107004.

\bibitem{dagh10}
M.~Daghofer, A.~Nicholson, A.~Moreo, E.~Dagotto, %{Three orbital model for the iron-based superconductors}, 
  Phys. Rev. B 81 (2010) 014511.

\bibitem{yu11}
R.~Yu, Q.~Si, %{Mott transition in multiorbital models for iron pnictides},
  Phys. Rev. B 84 (2011) 235115.

\bibitem{kane10}
E.~Kaneshita, T.~Tohyama, %{Spin and charge dynamics ruled by antiferromagnetic order in iron pnictide superconductors}, 
  Phys. Rev. B 82 (2010) 094441.

\bibitem{tham12}
V.~Thampy, J.~Kang, J.~A. Rodriguez-Rivera, W.~Bao, A.~T. Savici, J.~Hu, T.~J.
  Liu, B.~Qian, D.~Fobes, Z.~Q. Mao, C.~B. Fu, W.~C. Chen, Q.~Ye, R.~W. Erwin,
  T.~R. Gentile, Z.~Tesanovic, C.~Broholm, %{Friedel-Like Oscillations from Interstitial Iron in Superconducting ${\mathrm{Fe}}_{1+y}{\mathrm{Te}}_{0.62}{\mathrm{Se}}_{0.38}$}, 
  Phys. Rev. Lett. 108 (2012) 107002.

\bibitem{ewin11}
R.~A. Ewings, T.~G. Perring, J.~Gillett, S.~D. Das, S.~E. Sebastian, A.~E.
  Taylor, T.~Guidi, A.~T. Boothroyd, %{Itinerant spin excitations in {SrFe$_{2}$As$_{2}$} measured by inelastic neutron scattering}, 
  Phys. Rev. B 83 (2011) 214519.

\bibitem{erem10}
I.~Eremin, A.~V. Chubukov, %{Magnetic degeneracy and hidden metallicity of the spin-density-wave state in ferropnictides}, 
  Phys. Rev. B 81 (2010) 024511.

\bibitem{haul09}
K.~Haule, G.~Kotliar, %{Coherence--incoherence crossover in the normal state of iron oxypnictides and importance of Hund's rule coupling}, 
  New J. Phys. 11 (2009) 025021.

\bibitem{yin11}
Z.~P. Yin, K.~Haule, G.~Kotliar,
 % {{Magnetism and charge dynamics in iron pnictides}},
   Nat. Phys. 7 (2011) 294--297.

\bibitem{aich10}
M.~Aichhorn, S.~Biermann, T.~Miyake, A.~Georges, M.~Imada, %{Theoretical evidence for strong correlations and incoherent metallic state in FeSe},
  Phys. Rev. B 82 (2010) 064504.

\bibitem{park11b}
H.~Park, K.~Haule, G.~Kotliar, %{Magnetic Excitation Spectra in  ${\mathrm{BaFe}}_{2}{\mathrm{As}}_{2}$: A Two-Particle Approach within a  Combination of the Density Functional Theory and the Dynamical Mean-Field Theory Method}, 
  Phys. Rev. Lett. 107 (2011) 137007.

\bibitem{liu12}
M.~S. Liu, L.~W. Harriger, H.~Q. Luo, M.~Wang, R.~A. Ewings, T.~Guidi, H.~Park,
  K.~Haule, G.~Kotliar, S.~M. Hayden, P.~C. Dai, %{Nature of magnetic excitations in superconducting BaFe$_{1.9}$Ni$_{0.1}$As$_2$}, 
  Nat. Phys. 8 (2012) 376--381.

\bibitem{harr11}
L.~W. Harriger, H.~Q. Luo, M.~S. Liu, C.~Frost, J.~P. Hu, M.~R. Norman, P.~Dai,
  %{Nematic spin fluid in the tetragonal phase of BaFe${}_{2}$As${}_{2}$}, 
  Phys. Rev. B 84 (2011) 054544.

\bibitem{dial09}
S.~O. Diallo, V.~P. Antropov, T.~G. Perring, C.~Broholm, J.~J. Pulikkotil,
  N.~Ni, S.~L. Bud'ko, P.~C. Canfield, A.~Kreyssig, A.~I. Goldman, R.~J.
  McQueeney, %{Itinerant Magnetic Excitations in Antiferromagnetic CaFe$_2$As$_2$}, 
  Phys. Rev. Lett. 102 (2009) 187206.

\bibitem{wang11}
M.~Wang, X.~C. Wang, D.~L. Abernathy, L.~W. Harriger, H.~Q. Luo, Y.~Zhao, J.~W.
  Lynn, Q.~Q. Liu, C.~Q. Jin, C.~Fang, J.~Hu, P.~Dai, %{Antiferromagnetic spin excitations in single crystals of nonsuperconducting Li${}_{1-x}$FeAs}, 
  Phys. Rev. B 83 (2011) 220515.

\bibitem{zali11}
I.~A. Zaliznyak, Z.~Xu, J.~M. Tranquada, G.~Gu, A.~M. Tsvelik, M.~B. Stone,
 % {Unconventional temperature enhanced magnetism in iron telluride}, 
  Phys. Rev. Lett. 107 (2011) 216403.

\bibitem{joha09}
M.~D. Johannes, I.~I. Mazin, %{Microscopic origin of magnetism and magnetic interactions in ferropnictides}, 
  Phys. Rev. B 79 (2009) 220510.

\bibitem{yild09}
T.~Yildirim, %{Strong Coupling of the Fe-Spin State and the As-As Hybridization  in Iron-Pnictide Superconductors from First-Principle Calculations}, 
  Phys. Rev. Lett. 102 (2009) 037003.

\bibitem{yin11b}
Z.~P. Yin, K.~Haule, G.~Kotliar,
  %{{Kinetic frustration and the nature of the magnetic and paramagnetic states in iron pnictides and iron  chalcogenides}}, 
  Nat. Mater. 10 (2011) 932--935.

\bibitem{zhao09}
J.~Zhao, D.~T. Adroja, D.-X. Yao, R.~Bewley, S.~Li, X.~F. Wang, G.~Wu, X.~H.
  Chen, J.~Hu, P.~Dai, %Spin waves and magnetic exchange interactions in cafe2as2, 
  Nat. Phys. 5 (2009) 555--560.

\bibitem{krug09}
F.~Kruger, S.~Kumar, J.~Zaanen, J.~van~den Brink, %Spin-orbital frustrations and anomalous metallic state in iron-pnictide superconductors, 
  Phys. Rev. B 79 (2009) 054504.

\bibitem{lee09}
C.-C. Lee, W.-G. Yin, W.~Ku, %Ferro-orbital order and strong magnetic anisotropy in the parent compounds of iron-pnictide superconductors, 
  Phys. Rev. Lett. 103 (2009) 267001.

\bibitem{chen09a}
C.-C. Chen, B.~Moritz, J.~van~den Brink, T.~P. Devereaux, R.~R.~P. Singh,
  %Finite-temperature spin dynamics and phase transitions in spin-orbital models, 
  Phys. Rev. B 80 (2009) 180418(R).

\bibitem{lv09}
W.~Lv, J.~Wu, P.~Phillips, %Orbital ordering induces structural phase transition and the resistivity anomaly in iron pnictides, 
  Phys. Rev. B 80 (2009) 224506.

\bibitem{yi11}
M.~Yi, D.~Lu, J.-H. Chu, J.~G. Analytis, A.~P. Sorini, A.~F. Kemper, B.~Moritz,
  S.-K. Mo, R.~G. Moore, M.~Hashimoto, W.-S. Lee, Z.~Hussain, T.~P. Devereaux,
  I.~R. Fisher, Z.-X. Shen, %{Symmetry-breaking orbital anisotropy observed for  detwinned Ba(Fe$_{1-x}$Co$_x$)$_2$As$_2$ above the spin density wave transition}, 
  Proc. Natl. Acad. Sci. USA 108 (2011) 6878--6883.

\bibitem{naka11}
M.~Nakajima, T.~Liang, S.~Ishida, Y.~Tomioka, K.~Kihou, C.~H. Lee, A.~Iyo,
  H.~Eisaki, T.~Kakeshita, T.~Ito, S.~Uchida, %{Unprecedented anisotropic  metallic state in undoped iron arsenide BaFe2As2 revealed by optical spectroscopy}, 
  Proc. Natl. Acad. Sci. USA 108 (2011) 12238--12242.

\bibitem{zhao08}
J.~Zhao, Q.~Huang, C.~de~la Cruz, S.~Li, J.~W. Lynn, Y.~Chen, M.~A. Green,
  G.~F. Chen, G.~Li, Z.~Li, J.~L. Luo, N.~L. Wang, P.~Dai, %{Structural and magnetic phase diagram of CeFeAsO$_{1- x}$F$_x$ and its relation to high-temperature superconductivity}, 
  Nat. Mater. 7 (2008) 953--959.

\bibitem{chu09a}
J.-H. Chu, J.~G. Analytis, C.~Kucharczyk, I.~R. Fisher, %Determination of the  phase diagram of the electron-doped superconductor $ ba ( fe1-{}x cox ) 2 as2 $, 
  Phys. Rev. B 79 (2009) 014506.

\bibitem{rotu10}
C.~R. Rotundu, B.~Freelon, T.~R. Forrest, S.~D. Wilson, P.~N. Valdivia,
  G.~Pinuellas, A.~Kim, J.-W. Kim, Z.~Islam, E.~Bourret-Courchesne, N.~E.
  Phillips, R.~J. Birgeneau, %{Heat capacity study of BaFe$_{2}$As$_{2}$: Effects of annealing}, 
  Phys. Rev. B 82 (2010) 144525.

\bibitem{nand10}
S.~Nandi, M.~G. Kim, A.~Kreyssig, R.~M. Fernandes, D.~K. Pratt, A.~Thaler,
  N.~Ni, S.~L. Bud'ko, P.~C. Canfield, J.~Schmalian, R.~J. McQueeney, A.~I.
  Goldman, %Anomalous suppression of the orthorhombic lattice distortion in superconducting $ba(fe1-xcox)2as2$ single crystals, 
  Phys. Rev. Lett. 104 (2010) 057006.

\bibitem{ni10}
N.~Ni, A.~Thaler, J.~Q. Yan, A.~Kracher, E.~Colombier, S.~L. Bud'ko, P.~C.
  Canfield, S.~T. Hannahs, %{Temperature versus doping phase diagrams for  Ba(Fe$_{1-x}$TM$_{x}$)$_{2}$As$_{2}$ (TM =Ni,Cu,Cu/Co) single crystals},
  Phys. Rev. B 82 (2010) 024519.

\bibitem{kim11a}
M.~G. Kim, D.~K. Pratt, G.~E. Rustan, W.~Tian, J.~L. Zarestky, A.~Thaler, S.~L.
  Bud'ko, P.~C. Canfield, R.~J. McQueeney, A.~Kreyssig, A.~I. Goldman,
  %{Magnetic ordering and structural distortion in Ru-doped  BaFe${}_{2}$As${}_{2}$ single crystals studied by neutron and x-ray diffraction}, 
  Phys. Rev. B 83 (2011) 054514.

\bibitem{kim11b}
M.~G. Kim, R.~M. Fernandes, A.~Kreyssig, J.~W. Kim, A.~Thaler, S.~L. Bud'ko,
  P.~C. Canfield, R.~J. McQueeney, J.~Schmalian, A.~I. Goldman, %{Character of the structural and magnetic phase transitions in the parent and electron-doped BaFe${}_{2}$As${}_{2}$ compounds}, 
  Phys. Rev. B 83 (2011) 134522.

\bibitem{mart11}
K.~Marty, A.~D. Christianson, C.~H. Wang, M.~Matsuda, H.~Cao, L.~H. VanBebber,
  J.~L. Zarestky, D.~J. Singh, A.~S. Sefat, M.~D. Lumsden, %{Competing magnetic ground states in nonsuperconducting  Ba(${\mathrm{Fe}}_{1-x}{\mathrm{Cr}}_{x}$)${}_{2}$${\mathrm{As}}_{2}$ as seen via neutron diffraction}, 
  Phys. Rev. B 83 (2011) 060509(R).

\bibitem{bao09}
W.~Bao, Y.~Qiu, Q.~Huang, M.~A. Green, P.~Zajdel, M.~R. Fitzsimmons,
  M.~Zhernenkov, S.~Chang, M.~Fang, B.~Qian, E.~K. Vehstedt, J.~Yang, H.~M.
  Pham, L.~Spinu, Z.~Q. Mao, %Tunable (delta pi, delta pi)-type antiferromagnetic order in alpha-fe(te,se) superconductors, 
  Phys. Rev. Lett. 102 (2009) 247001.

\bibitem{rodr11}
E.~E. Rodriguez, C.~Stock, P.~Zajdel, K.~L. Krycka, C.~F. Majkrzak, P.~Zavalij,
  M.~A. Green, %{Magnetic-crystallographic phase diagram of the superconducting parent compound Fe${}_{1+x}$Te}, 
  Phys. Rev. B 84 (2011) 064403.

\bibitem{stoc11}
C.~Stock, E.~E. Rodriguez, M.~A. Green, P.~Zavalij, J.~A. Rodriguez-Rivera,
 % {Interstitial iron tuning of the spin fluctuations in the nonsuperconducting parent phase ${\mathbf{Fe}}_{1+x}\mathbf{Te}$}, 
  Phys. Rev. B 84 (2011) 045124.

\bibitem{li09a}
S.~Li, C.~de~la Cruz, Q.~Huang, Y.~Chen, J.~W. Lynn, J.~Hu, Y.-L. Huang, F.-C.
  Hsu, K.-W. Yeh, M.-K. Wu, P.~Dai, %First-order magnetic and structural phase transitions in fe[sub 1 + y]se[sub x]te[sub 1 - x], 
  Phys. Rev. B 79 (2009) 054503.

\bibitem{mart10b}
A.~Martinelli, A.~Palenzona, M.~Tropeano, C.~Ferdeghini, M.~Putti, M.~R.
  Cimberle, T.~D. Nguyen, M.~Affronte, C.~Ritter, %From antiferromagnetism to superconductivity in $ fe1+y te1-{}x sex $ $ ( \le{}x\le{}0.20 ) $ : Neutron powder diffraction analysis, 
  Phys. Rev. B 81 (2010) 094115.

\bibitem{chu12}
J.-H. Chu, H.-H. Kuo, J.~G. Analytis, I.~R. Fisher, %{Divergent Nematic Susceptibility in an Iron Arsenide Superconductor}, 
  Science 337 (2012) 710--712.

\bibitem{yosh12}
M.~Yoshizawa, D.~Kimura, T.~Chiba, S.~Simayi, Y.~Nakanishi, K.~Kihou, C.-H.
  Lee, A.~Iyo, H.~Eisaki, M.~Nakajima, S. Uchida, %{Structural Quantum Criticality and Superconductivity in Iron-Based Superconductor Ba(Fe$_{1-x}$Co$_{x}$)$_{2}$As$_{2}$}, 
  J. Phys. Soc. Jpn. 81 (2012) 024604.

\bibitem{wyso11}
A.~L. Wysocki, K.~D. Belashchenko, V.~P. Antropov, %{Consistent model of magnetism in ferropnictides}, 
  Nat. Phys. 7 (2011) 485--489.

\bibitem{mazi09}
I.~I. Mazin, M.~D. Johannes, %{A key role for unusual spin dynamics in ferropnictides}, 
  Nat. Phys. 5 (2009) 141--145.

\bibitem{moon10}
C.-Y. Moon, H.~J. Choi, %{Chalcogen-Height Dependent Magnetic Interactions and Magnetic Order Switching in FeSe$_x$Te$_{1-x}$}, 
  Phys. Rev. Lett. 104 (2010) 057003.

\bibitem{kuro08}
K.~Kuroki, S.~Onari, R.~Arita, H.~Usui, Y.~Tanaka, H.~Kontani, H.~Aoki,
 % Unconventional pairing originating from the disconnected fermi surfaces of superconducting lafeaso[sub 1-x]f[sub x], 
  Phys. Rev. Lett. 101 (2008) 087004.

\bibitem{dagh08}
M.~Daghofer, A.~Moreo, J.~A. Riera, E.~Arrigoni, D.~J. Scalapino, E.~Dagotto,
  %{Model for the Magnetic Order and Pairing Channels in Fe Pnictide Superconductors}, 
  Phys. Rev. Lett. 101 (2008) 237004.

\bibitem{knol10}
J.~Knolle, I.~Eremin, A.~V. Chubukov, R.~Moessner, %{Theory of itinerant magnetic excitations in the spin-density-wave phase of iron-based superconductors}, 
  Phys. Rev. B 81 (2010) 140506.

\bibitem{lee10b}
Y.~Lee, D.~Vaknin, H.~Li, W.~Tian, J.~L. Zarestky, N.~Ni, S.~L. Bud'ko, P.~C.
  Canfield, R.~J. McQueeney, %B.~N. Harmon, {Magnetic form factor of iron in SrFe$_{2}$As$_{2}$}, 
  Phys. Rev. B 81 (2010) 060406.

\bibitem{ratc10}
W.~Ratcliff, P.~A. Kienzle, J.~W. Lynn, S.~Li, P.~Dai, G.~F. Chen, N.~L. Wang,
  %{Magnetic form factor of SrFe$_{2}$As$_{2}$: Neutron diffraction measurements}, 
  Phys. Rev. B 81 (2010) 140502.

\bibitem{brow10}
P.~J. Brown, T.~Chatterji, A.~Stunault, Y.~Su, Y.~Xiao, R.~Mittal,
  T.~Br\"uckel, T.~Wolf, P.~Adelmann, %{Magnetization distribution in the tetragonal phase of BaFe$_{2}$As$_{2}$}, 
  Phys. Rev. B 82 (2010) 024421.

\bibitem{prok11}
K.~Proke{\v s}, A.~Gukasov, D.~N. Argyriou, S.~L. Bud'ko, P.~C. Canfield,
  A.~Kreyssig, A.~I. Goldman, %{Magnetization distribution in the tetragonal Ba(Fe 1−x Co x ) 2 As 2 , x=0.066 probed by polarized neutron diffraction},
  Europhys. Lett. 93 (2011) 32001.

\bibitem{lest11}
C.~Lester, J.-H. Chu, J.~G. Analytis, A.~Stunault, I.~R. Fisher, S.~M. Hayden,
  %{Polarized neutron diffraction study of the field-induced magnetization in  the normal and superconducting states of Ba(Fe$_{1-x}$Co$_{x}$)$_{2}$As$_{2}$ ($x=0.65$)}, 
  Phys. Rev. B 84 (2011) 134514.

\bibitem{luet09}
H.~Luetkens, H.~H. Klauss, M.~Kraken, F.~J. Litterst, T.~Dellmann,
  R.~Klingeler, C.~Hess, R.~Khasanov, A.~Amato, C.~Baines, M.~Kosmala, O.~J.
  Schumann, M.~Braden, J.~Hamann-Borrero, N.~Leps, A.~Kondrat, G.~Behr,
  J.~Werner, B.~Buchner, %{The electronic phase diagram of the LaO1-xFxFeAs superconductor}, 
  Nat. Mater. 8 (2009) 305--309.

\bibitem{chen09b}
H.~Chen, Y.~Ren, Y.~Qiu, W.~Bao, R.~H. Liu, G.~Wu, T.~Wu, Y.~L. Xie, X.~F.
  Wang, Q.~Huang, X.~H. Chen, %{Coexistence of the spin-density wave and superconductivity in Ba1-xKxFe2As2}, 
  Europhys. Lett. 85 (2009) 17006.

\bibitem{drew09}
A.~J. Drew, C.~Niedermayer, P.~J. Baker, F.~L. Pratt, S.~J. Blundell,
  T.~Lancaster, R.~H. Liu, G.~Wu, X.~H. Chen, I.~Watanabe, V.~K. Malik,
  A.~Dubroka, M.~Rossle, K.~W. Kim, C.~Baines, C.~Bernhard, %{Coexistence of  static magnetism and superconductivity in SmFeAsO1-xFx as revealed by muon spin rotation}, 
  Nat. Mater. 8 (2009) 310--314.

\bibitem{drew08}
A.~J. Drew, F.~L. Pratt, T.~Lancaster, S.~J. Blundell, P.~J. Baker, R.~H. Liu,
  G.~Wu, X.~H. Chen, I.~Watanabe, V.~K. Malik, A.~Dubroka, K.~W. Kim,
  M.~R\"ossle, C.~Bernhard, %{Coexistence of Magnetic Fluctuations and Superconductivity in the Pnictide High Temperature Superconductor ${\mathrm{SmFeAsO}}_{1-x}{\mathrm{F}}_{x}$ Measured by Muon Spin Rotation},
  Phys. Rev. Lett. 101 (2008) 097010.

\bibitem{lums10}
M.~D. Lumsden, A.~D. Christianson, E.~A. Goremychkin, S.~E. Nagler, H.~A. Mook,
  M.~B. Stone, D.~L. Abernathy, T.~Guidi, G.~J. MacDougall, C.~de~la Cruz,
  A.~S. Sefat, M.~A. McGuire, B.~C. Sales, D.~Mandrus, %Evolution of spin excitations into the superconducting state in fete1-xsex, 
  Nat. Phys. 6 (2010) 182--186.

\bibitem{sato74}
H.~Sato, K.~Maki, Int. J. Magn. 6 (1974) 183.

\bibitem{noak90}
D.~R. Noakes, T.~M. Holden, E.~Fawcett, P.~C. de~Camargo, %{Critical scattering in Cr+0.2 at.\%\ V and in chromium}, 
  Phys. Rev. Lett. 65 (1990) 369--372.

\bibitem{li10a}
H.-F. Li, C.~Broholm, D.~Vaknin, R.~M. Fernandes, D.~L. Abernathy, M.~B. Stone,
  D.~K. Pratt, W.~Tian, Y.~Qiu, N.~Ni, S.~O. Diallo, J.~L. Zarestky, S.~L.
  Bud'ko, P.~C. Canfield, R.~J. McQueeney, %{Anisotropic and quasipropagating spin excitations in superconducting Ba(Fe$_{0.926}$Co$_{0.074}$)$_{2}$As$_{2}$}, 
  Phys. Rev. B 82 (2010) 140503.

\bibitem{lest10}
C.~Lester, J.-H. Chu, J.~G. Analytis, T.~G. Perring, I.~R. Fisher, S.~M.
  Hayden, %{Dispersive spin fluctuations in the nearly optimally doped superconductor Ba(Fe$_{1-x}$Co$_{x}$)$_{2}$As$_{2}$ $(x=0.065)$}, 
  Phys. Rev. B 81 (2010) 064505.

\bibitem{chri08}
A.~D. Christianson, E.~A. Goremychkin, R.~Osborn, S.~Rosenkranz, M.~D. Lumsden,
  C.~D. Malliakas, I.~S. Todorov, H.~Claus, D.~Y. Chung, M.~G. Kanatzidis,
  R.~I. Bewley, T.~Guidi, %{Unconventional superconductivity in {Ba$_{0.6}$K$_{0.4}$Fe$_2$As$_2$} from inelastic neutron scattering}, 
  Nature 456 (2008) 930--932.

\bibitem{cast11}
J.~P. Castellan, S.~Rosenkranz, E.~A. Goremychkin, D.~Y. Chung, I.~S. Todorov,
  M.~G. Kanatzidis, I.~Eremin, J.~Knolle, A.~V. Chubukov, S.~Maiti, M.~R.
  Norman, F.~Weber, H.~Claus, T.~Guidi, R.~I. Bewley, R.~Osborn, %{Effect of Fermi surface nesting on resonant spin excitations in Ba$_{1-x}$K$_x$Fe$_2$As$_2$}, 
  Phys. Rev. Lett. 107 (2011) 177003.

\bibitem{lums09}
M.~D. Lumsden, A.~D. Christianson, D.~Parshall, M.~B. Stone, S.~E. Nagler,
  G.~J. MacDougall, H.~A. Mook, K.~Lokshin, T.~Egami, D.~L. Abernathy, E.~A.
  Goremychkin, R.~Osborn, M.~A. McGuire, A.~S. Sefat, R.~Jin, B.~C. Sales,
  D.~Mandrus, %{Two-dimensional resonant magnetic excitation in {BaFe$_{1.84}$Co$_{0.16}$As$_2$}}, 
  Phys. Rev. Lett. 102 (2009) 107005.

\bibitem{inos10}
D.~S. Inosov, J.~T. Park, P.~Bourges, D.~L. Sun, Y.~Sidis, A.~Schneidewind,
  K.~Hradil, D.~Haug, C.~T. Lin, B.~Keimer, V.~Hinkov, %{Normal-state spin dynamics and temperature-dependent spin-resonance energy in optimally doped BaFe1.85Co0.15As2}, 
  Nat. Phys. 6 (2010) 178--181.

\bibitem{chi09}
S.~Chi, A.~Schneidewind, J.~Zhao, L.~W. Harriger, L.~Li, Y.~Luo, G.~Cao, Z.~A.
  Xu, M.~Loewenhaupt, J.~Hu, P.~Dai, %{Inelastic Neutron-Scattering Measurements of a Three-Dimensional Spin Resonance in the {FeAs}-Based  {BaFe$_{1.9}$Ni$_{0.1}$As$_2$} Superconductor}, 
  Phys. Rev. Lett. 102 (2009) 107006.

\bibitem{li09c}
S.~Li, Y.~Chen, S.~Chang, J.~W. Lynn, L.~Li, Y.~Luo, G.~Cao, Z.~Xu, P.~Dai,
 % {Spin gap and magnetic resonance in superconducting BaFe$_{1.9}$Ni$_{0.1}$As$_2$}, 
  Phys. Rev. B 79 (2009) 174527.

\bibitem{qiu09}
Y.~Qiu, W.~Bao, Y.~Zhao, C.~Broholm, V.~Stanev, Z.~Tesanovic, Y.~C. Gasparovic,
  S.~Chang, J.~Hu, B.~Qian, M.~Fang, Z.~Mao, %{Spin Gap and Resonance at the Nesting Wave Vector in Superconducting FeSe$_{0.4}$Te$_{0.6}$}, 
  Phys. Rev. Lett. 103 (2009) 067008.

\bibitem{mook10}
H.~A. Mook, M.~D. Lumsden, A.~D. Christianson, S.~E. Nagler, B.~C. Sales,
  R.~Jin, M.~A. McGuire, A.~S. Sefat, D.~Mandrus, T.~Egami, C.~dela Cruz,
  %Unusual relationship between magnetism and superconductivity in $fete_{0.5}se_{0.5}$, 
  Phys. Rev. Lett. 104 (2010) 187002.

\bibitem{mart12}
K.~Marty, A.~D. Christianson, A.~M. dos Santos, B.~Sipos, K.~Matsubayashi,
  Y.~Uwatoko, J.~A. Fernandez-Baca, C.~A. Tulk, T.~A. Maier, B.~C. Sales, M.~D.
  Lumsden, %{Effect of pressure on the neutron spin resonance in the unconventional superconductor FeTe${}_{0.6}$Se${}_{0.4}$}, 
  Phys. Rev. B 86 (2012) 220509.

\bibitem{chub08}
A.~V. Chubukov, D.~V. Efremov, I.~Eremin, %{Magnetism, superconductivity, and pairing symmetry in iron-based superconductors}, 
  Phys. Rev. B 78 (2008) 134512.

\bibitem{maie09}
T.~A. Maier, S.~Graser, D.~J. Scalapino, P.~Hirschfeld, %{Neutron scattering resonance and the iron-pnictide superconducting gap}, 
  Phys. Rev. B 79 (2009) 134520.

\bibitem{onar11}
S.~Onari, H.~Kontani, %{Neutron inelastic scattering peak by dissipationless mechanism in the ${s}_{++}$-wave state in iron-based superconductors}, 
  Phys. Rev. B 84 (2011) 144518.

\bibitem{lee10}
S.-H. Lee, G.~Xu, W.~Ku, J.~S. Wen, C.~C. Lee, N.~Katayama, Z.~J. Xu, S.~Ji,
  Z.~W. Lin, G.~D. Gu, H.-B. Yang, P.~D. Johnson, Z.-H. Pan, T.~Valla,
  M.~Fujita, T.~J. Sato, S.~Chang, K.~Yamada, J.~M. Tranquada, %{Coupling of spin and orbital excitations in the iron-based superconductor FeSe$_{0.5}$Te$_{0.5}$}, 
  Phys. Rev. B 81 (2010) 220502.

\bibitem{kont10}
H.~Kontani, S.~Onari, %{Orbital-Fluctuation-Mediated Superconductivity in Iron Pnictides: Analysis of the Five-Orbital Hubbard-Holstein Model}, 
  Phys. Rev. Lett. 104 (2010) 157001.

\bibitem{wen10}
J.~Wen, G.~Xu, Z.~Xu, Z.~W. Lin, Q.~Li, Y.~Chen, S.~Chi, G.~Gu, J.~M.
  Tranquada, %Effect of magnetic field on the spin resonance in $fete_{0.5}se_{0.5}$ as seen via inelastic neutron scattering, 
  Phys. Rev. B 81 (2010) 100513(R).

\bibitem{zhao10}
J.~Zhao, L.-P. Regnault, C.~Zhang, M.~Wang, Z.~Li, F.~Zhou, Z.~Zhao, C.~Fang,
  J.~Hu, P.~Dai, %{Neutron spin resonance as a probe of the superconducting energy gap of BaFe$_{1.9}$Ni$_{0.1}$As$_{2}$ superconductors}, 
  Phys. Rev. B 81 (2010) 180505.

\bibitem{bao10}
W.~Bao, A.~T. Savici, G.~E. Granroth, C.~Broholm, K.~Habicht, Y.~Qiu, J.~Hu,
  T.~Liu, Z.~Mao, {A Triplet Resonance in Superconducting
  FeSe$_{0.4}$Te$_{0.6}$}, arXiv:1002.1617v1.

\bibitem{xu12a}
Z.~Xu, J.~Wen, Y.~Zhao, M.~Matsuda, W.~Ku, X.~Liu, G.~Gu, D.-H. Lee, R.~J.
  Birgeneau, J.~M. Tranquada, G.~Xu, %{Temperature-Dependent Transformation of the Magnetic Excitation Spectrum on Approaching Superconductivity in  ${\mathrm{Fe}}_{1+y-x}(\mathrm{Ni}/\mathrm{Cu}{)}_{x}{\mathrm{Te}}_{0.5}{\mathrm{Se}}_{0.5}$},
  Phys. Rev. Lett. 109 (2012) 227002.

\bibitem{arha12}
H.~Z. Arham, C.~R. Hunt, W.~K. Park, J.~Gillett, S.~D. Das, S.~E. Sebastian,
  Z.~J. Xu, J.~S. Wen, Z.~W. Lin, Q.~Li, G.~Gu, A.~Thaler, S.~Ran, S.~L.
  Bud'ko, P.~C. Canfield, D.~Y. Chung, M.~G. Kanatzidis, L.~H. Greene,
  %{Detection of orbital fluctuations above the structural transition temperature in the iron pnictides and chalcogenides}, 
  Phys. Rev. B 85 (2012) 214515.

\bibitem{tsyr12}
N.~Tsyrulin, R.~Viennois, E.~Giannini, M.~Boehm, M.~Jimenez-Ruiz, A.~A. Omrani,
  B.~D. Piazza, H.~M. R{\o}nnow, %{Magnetic hourglass dispersion and its relation to high-temperature superconductivity in iron-tuned Fe$_{1+y}$Te$_{0.7}$Se$_{0.3}$}, 
  New J. Phys. 14 (2012) 073025.

\bibitem{pall09}
I.~Pallecchi, G.~Lamura, M.~Tropeano, M.~Putti, R.~Viennois, E.~Giannini,
  D.~Van~der Marel, %{Seebeck effect in Fe$_{1+x}$Te$_{1-y}$Se$_{y}$ single  crystals}, 
  Phys. Rev. B 80 (2009) 214511.

\bibitem{home10}
C.~C. Homes, A.~Akrap, J.~S. Wen, Z.~J. Xu, Z.~W. Lin, Q.~Li, G.~D. Gu,
  %{Electronic correlations and unusual superconducting response in the optical  properties of the iron chalcogenide {FeTe$_{0.55}$Se$_{0.45}$}}, 
  Phys. Rev. B 81 (2010) 180508.

\bibitem{moon11}
S.~J. Moon, C.~C. Homes, A.~Akrap, Z.~J. Xu, J.~S. Wen, Z.~W. Lin, Q.~Li, G.~D.
  Gu, D.~N. Basov, %{Incoherent $c$-Axis Interplane Response of the Iron Chalcogenide ${\mathrm{FeTe}}_{0.55}{\mathrm{Se}}_{0.45}$ Superconductor from Infrared Spectroscopy}, 
  Phys. Rev. Lett. 106 (2011) 217001.

\end{thebibliography}

\end{document}